\pdfoutput=1
\documentclass{aa}  

\usepackage{graphicx}
\usepackage{txfonts}
 \usepackage{textcomp}
\usepackage{natbib,twoopt}
\bibpunct{(}{)}{;}{a}{}{,} %% natbib format for A&A and ApJ
\makeatletter
\newcommandtwoopt{\citeads}[3][][]{\href{http://adsabs.harvard.edu/abs/#3}%
{\def\hyper@linkstart##1##2{}%
\let\hyper@linkend\@empty\citealp[#1][#2]{#3}}}
\newcommandtwoopt{\citepads}[3][][]{\href{http://adsabs.harvard.edu/abs/#3}%
{\def\hyper@linkstart##1##2{}%
\let\hyper@linkend\@empty\citep[#1][#2]{#3}}}
\newcommandtwoopt{\citetads}[3][][]{\href{http://adsabs.harvard.edu/abs/#3}%
{\def\hyper@linkstart##1##2{}%
\let\hyper@linkend\@empty\citet[#1][#2]{#3}}}
\newcommandtwoopt{\citeyearads}[3][][]%
{\href{http://adsabs.harvard.edu/abs/#3}
{\def\hyper@linkstart##1##2{}%
\let\hyper@linkend\@empty\citeyear[#1][#2]{#3}}}
\makeatother

\begin{document}

   \title{Probing embedded star clusters in the HII complex NGC\,6357 with 
   VVV\thanks{Based on observations taken with the ESO VISTA 
   Public Survey VVV, Programme ID 179.B-2002 and data from the 2MASS 
   VizieR Catalog II/246.}}

   \author{E. F. Lima,
          \inst{1,}
          \inst{2}
          E. Bica,
          \inst{1}
          C. Bonatto
          \inst{1}
          \and
          R. K. Saito\inst{3}
          }

   \institute{Universidade Federal do Rio Grande do Sul, Departamento de Astronomia\\               
              CP 15051, Porto Alegre 91501-970, Brazil\\
              \email{eliade.lima@ufrgs.br, bica@if.ufrgs.br, charles@if.ufrgs.br}\\
         \and     
            Departamento de F\'\i sica, CCNE, Universidade Federal de Santa Maria\\\
             97105-900, Santa Maria, RS, Brazil\\
         \and
            Departamento de F\'\i sica, Universidade Federal de Sergipe\\            
            Av. Marechal Rondon s/n, 49100-000, São Crist\'ov\~ao, Brazil\\
             }

   \date{Received xxxxxxx; accepted xxxxxx}

%%%%%%%%%%abstract%%%%%%%%%%%%%%%%%%%%%%%%%%%%%%%%%%%%%%%%%%%%%%%%%%%%%%%%%% 
 
  \abstract
  % context  
   {NGC\,6357 is an active star-forming region located in the Sagittarius
   arm displaying several star clusters, which makes it a very interesting target 
   to investigate star formation and early cluster evolution.}
  % aims
   {We explore NGC\,6357 with the ``VISTA Variables in the V\'\i a L\'actea'' 
   (VVV) photometry of seven embedded clusters (ECs), and one open cluster (OC) projected 
   in the outskirts of the complex.}
  % methods 
   {Photometric and structural properties (age, reddening, distance, core and 
   total radii) of the star clusters are derived. VVV saturated stars are replaced 
   by their 2MASS counterparts. Field-decontaminated VVV photometry is used to analyse 
   Colour-Magnitude Diagrams (CMDs), stellar radial density profiles (RDPs) and 
   determine astrophysical parameters.}
  % results 
   {We report the discovery of four ECs and one intermediate-age cluster in the complex
   area. We derive a revised distance estimate for NGC\,6357 of $1.78\pm0.1$\,kpc based on the 
   cluster CMD morphologies. Among the ECs, one contains the binary star the WR\,93, while
   the remaining ones are dominated by pre-main sequence (PMS) stars, young-stellar objects 
   (YSO) and/or and have a developed main sequence. These features reflect a significant age spread
   among the clusters. Evidence is found that the relatively populous cluster Pismis\,24 hosts 
   two subclusters.  }
  % conclusions 
   {}

                
   \titlerunning{Star clusters in NGC\,6357}

   \maketitle
%
%%%%%%%%%%%%%%%%%%%%%%%%%%%Introduction%%%%%%%%%%%%%%%%%%%%%%%%%%%%%%%%%%%%%%%
\section{Introduction}
\label{introd}

   Star-forming complexes are, in general, major building blocks of the
   large-scale structure of galaxies and important sites to study how
   massive stars form \citep{Russeil_2010}. In particular, Galactic embedded and
   open clusters are excellent probes of the structure and evolution
   of the disk and spiral arms \citep{Carvalho_2008, 2003ARA&A..41...57L, 1995ARA&A..33..381F}.
   Embedded clusters (ECs) can be partially or fully immersed in embryonic
   molecular clouds and HII regions. According to \citet{1989ApJS...70..731L}, all clusters
   younger than $\sim5$ Myr are connected at least to one large molecular cloud or HII region.
   
   Recently, our group studied the stellar content of the Sh2--132 HII 
   region, a star-forming complex hosting at least 4 ECs 
   and presenting evidence of triggered star formation and hierarchical 
   structuring \citep{2010MNRAS.407..133S}. Sequential star formation in giant molecular clouds was also  
   studied by \citet{2011MNRAS.416.1522C}.
   Observationally, low-mass star clusters younger than about 10 Myr 
   present an underpopulated, developing main sequence (MS) and a more 
   populous feature of pre-main sequence (PMS) stars 
   \citep{2009MNRAS.392..483B}.
   Studies of very young star clusters hosting PMS and MS stars have produced
   well-defined CMDs, RDPs and mass functions \citep{2009MNRAS.397.1915B}. 
   Recently additional tools were developed by our group allowing to obtain reliable fundamental 
   parameters of early--cluster phases \citep{2012MNRAS.420..352B, 2012A&A...540A.137B}.
   The present paper focuses on the ECs in the NGC\,6357 complex.
   
   NGC\,6357 ($\equiv$ W 22 $\equiv$ RCW 131 $\equiv$ Sh2--11) is a large HII region complex
   that consists of a shell of about 60 $\times$ 40 arcmin$^2$,
   bright optical nebulosities in different  evolutionary stages,
   OB stars belonging to the populous open cluster Pismis\,24 and
   YSOs candidates \citep{Felli_1990, Bohigas_2004, Wang_2007,
   Russeil_2010, Fang_2012}. \citet{1986A&A...170...97P} showed
   that the whole HII complex is an active area of recent and on-going star formation. 
   The shell has been interpreted as an ionized gas bubble created by the strong
   winds of the current massive stars in Pismis\,24
   or by a previous generation \citep{Lortet_1984, 
   Bohigas_2004, Wang_2007}. The total amount of molecular
   gas related to the large shell was estimated by \citet{Cappa_2011}
   as 1.4 $\times$ 10$^5$ M$_{\odot}$.
   
   Early optical studies of NGC\,6357 and Pismis\,24 revealed $\sim$20 O-type and early
   B-type stars \citep{Moffat_1973, Neckel_1978, Neckel_1984, Lortet_1984}, including a binary
   system (HD\,157504) composed of a WC7 Wolf-Rayet star (WR 93) and an O7-9 star \citep{Van_der_Hucht_01}.
   Two of the cluster members, namely Pismis\,24-1 (HDE\,319718) and Pismis\,24-17, 
   were recently classified as spectral type O3.5, some of the brightest and bluest 
   stars known \citep{Massey_2001, Walborn_2002}. The total to selective extinction ratio (R$_V$)
   towards NGC\,6357 appears to be about 3.5 \citep{Russeil_2012, Bohigas_2004}. 
   %We adopt R$_V$ = 3.1 \citep{Wegner_1994, Whittet_80, Rieke_85}.
   
   A wide range of distances (1.1 -- 2.6 kpc) has been derived to NGC\,6357.
   This is usually estimated from the distance of Pismis\,24. The most 
   recent determination is that of \citet{Fang_2012},
   who give 1.7$\pm$0.2 kpc. The kinematic distance is d$_{\odot}$=1.0$\pm$2.3 kpc 
   \citep{Wilson_1970}. \citet{Neckel_1978} obtain d$_{\odot}$=1.74$\pm$0.31 kpc and
   \citet{Massey_2001} find a distance of 2.56$\pm$0.10 kpc for Pismis\,24. 
   \citet{Conti_1990} and \citet{Van_der_Hucht_01} derived for WR 93 d$_{\odot}$=1.1 kpc and
   d$_{\odot}$=1.74 kpc, respectively.
   
   Optical, radio continuum, and near- and mid-IR images of NGC\,6357
   \citep{Cappa_2011} indicate the giant nature of this complex and its evolved 
   character, which suggest a suitable laboratory for early dynamical and hydrodynamical
   evolution. The number of ECs in a given star-forming region
   is fundamental to study the gas expulsion and dynamical evolution effects
   \citep{Carvalho_2008}. Using the recent VVV JHK$_S$ data \citep{Minniti_2010}, we analyse in detail in the present paper
   the stellar clusters detected in NGC\,6357. We employ a field-decontamination algorithm
   \citep{2007MNRAS.377.1301B} adapted to VVV photometry to analyse CMDs, RDPs and determine the
   astrophysical parameters.
   The present analysis of the ECs shows structural variations as 
   well as morphology diversity of MS and PMS evolutionary sequences in CMDs. 
   Based on this, we infer the star formation history in the complex.
   
   This paper is organised as follows. In Sect.~\ref{sc_6357} the stellar cluster sample is described.
   In Sect.~\ref{VVV_data} we provide details on the VVV data adopted. In Sects.~\ref{SCA} and \ref{structure}, the cluster
   photometric and structural analyses are carried out. In Sect.~\ref{concl} we discuss the results and give conclusions.

  %%%%%%%%%%%%%%%%%%%%%%%%%%%%%%%%%%%%%%%%%%%%%%%%%%%%%%%%%%%%%%%%%
 \section{Stellar clusters in NGC\,6357}
 \label{sc_6357}
 
   \begin{figure*}
   \centering
   \includegraphics[scale=0.43]{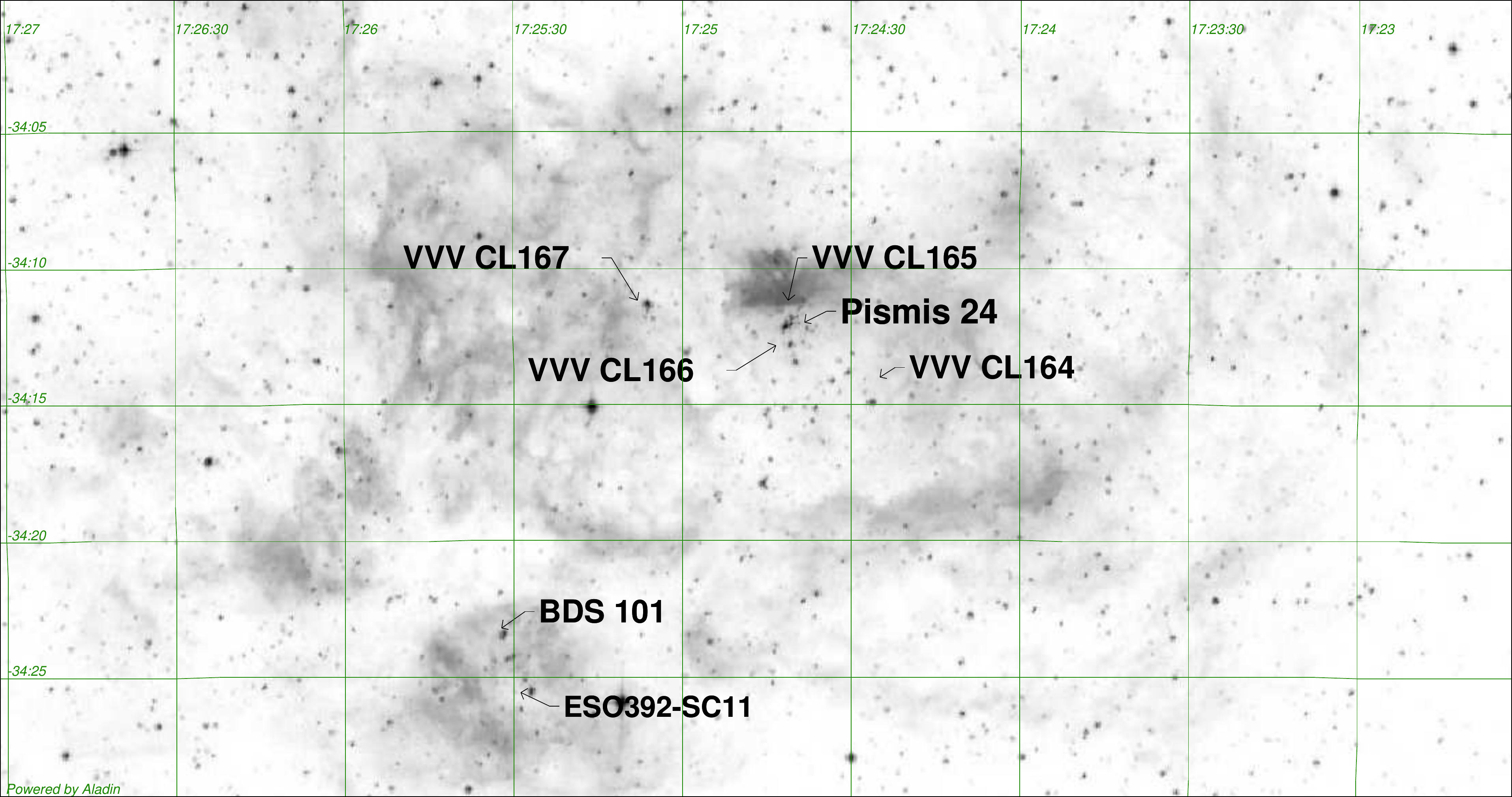}
   \caption{55$'$ $\times$ 30$'$ DSS R image of the emission nebula NGC\,6357. We indicate the present sample of star clusters. }
   \label{global}
  \end{figure*}
 
   \begin{figure*}
   \centering
   {\includegraphics[scale=0.39]{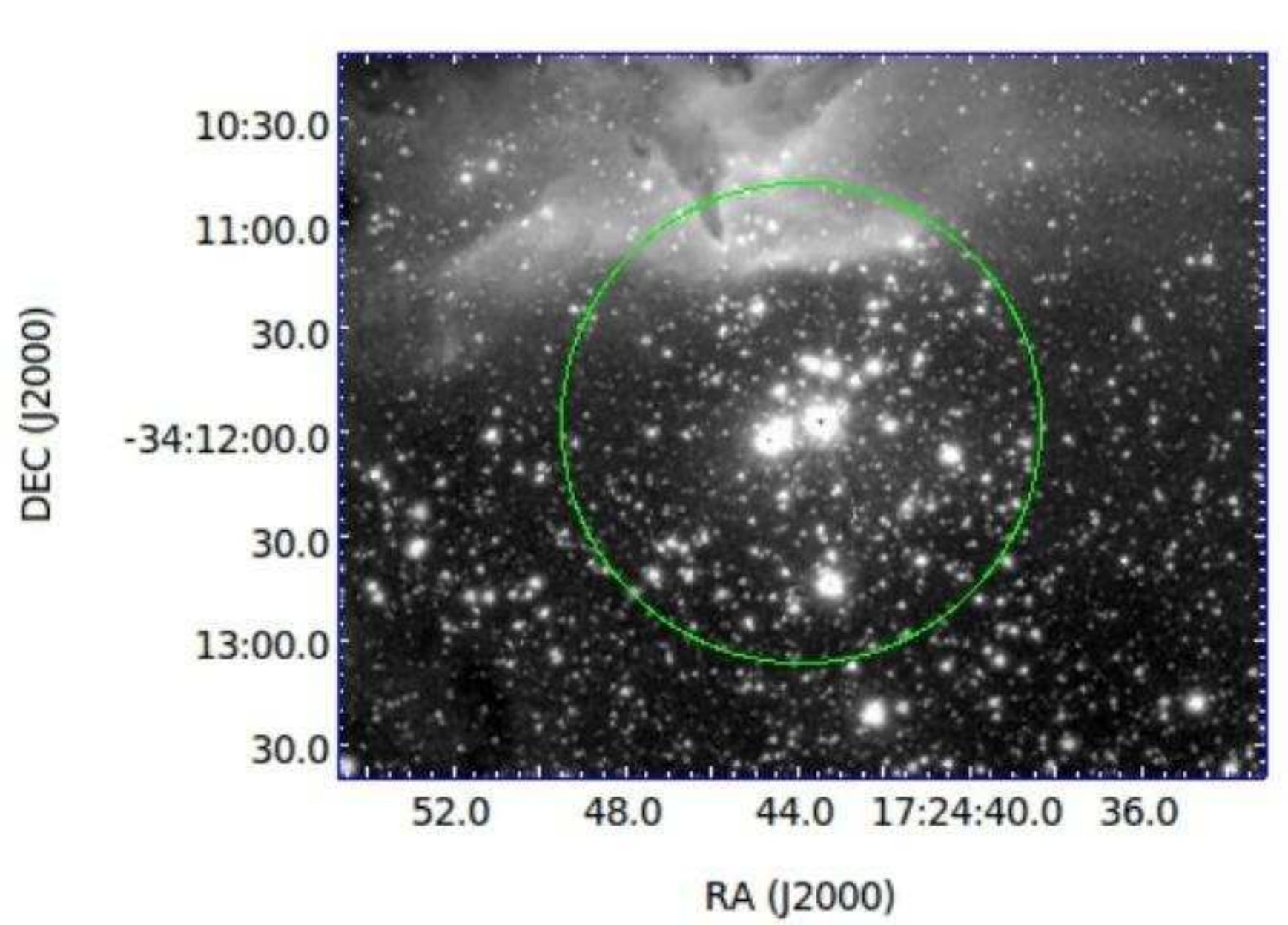}
   \includegraphics[scale=0.44]{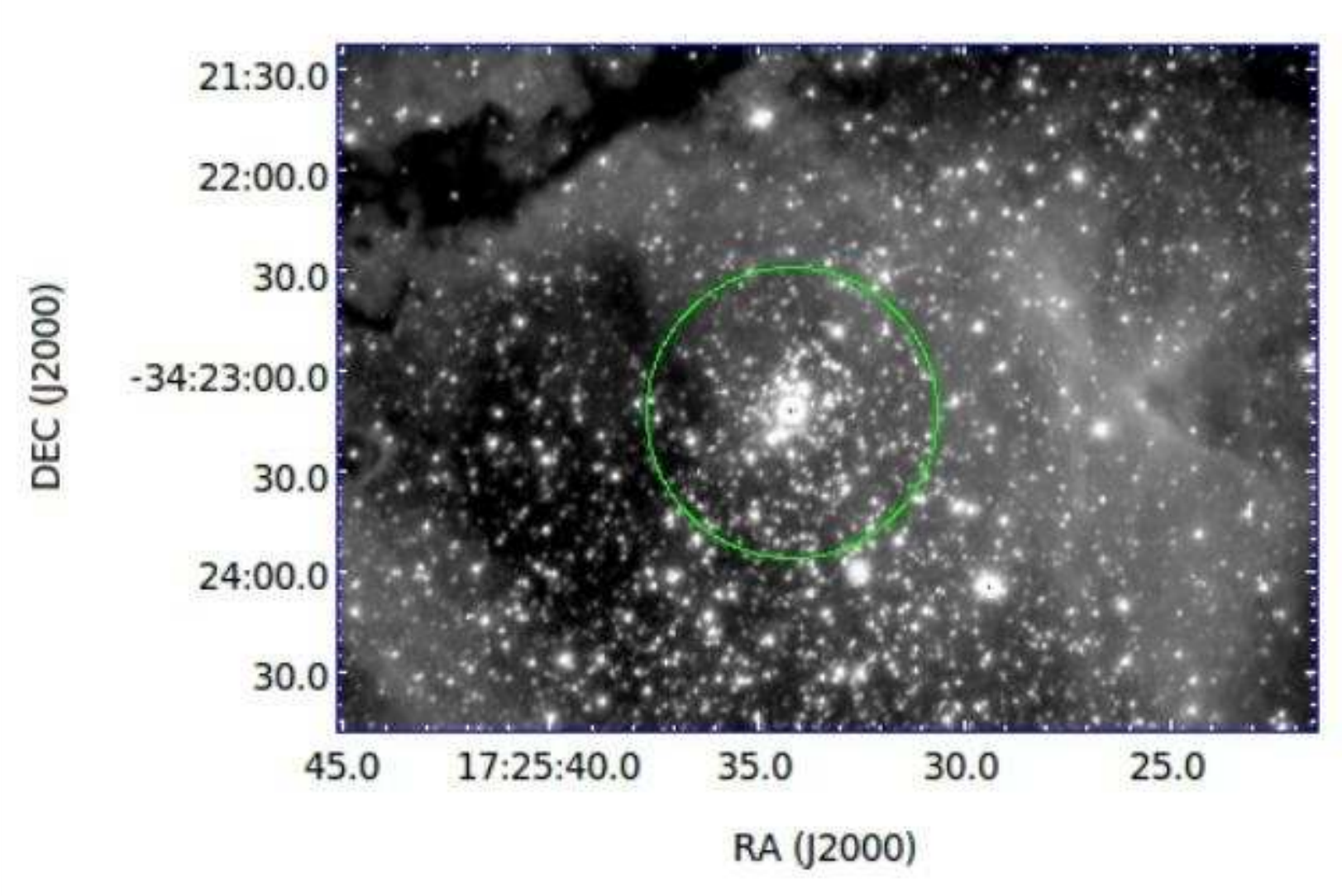}   
   \includegraphics[scale=0.43]{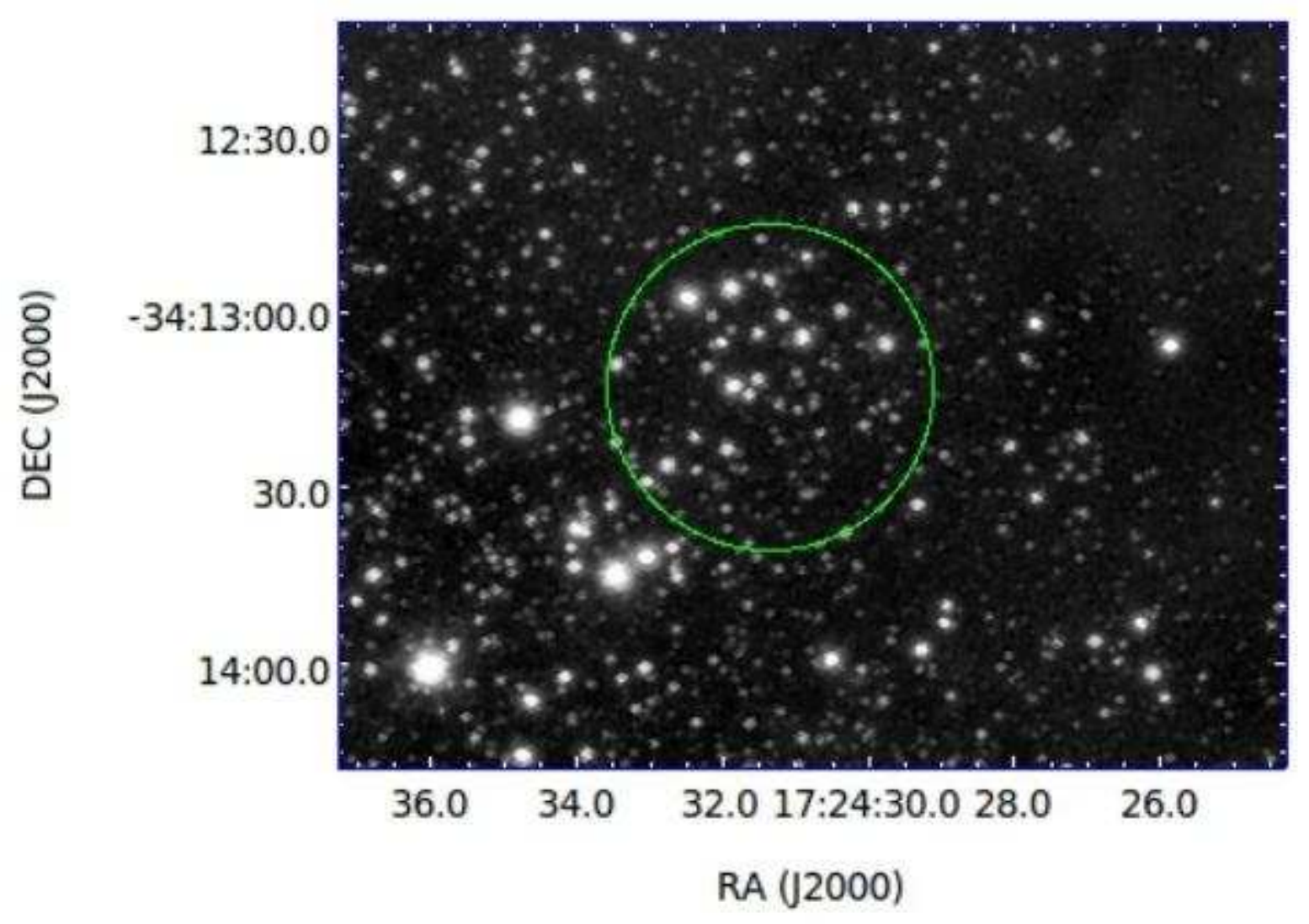}
   \includegraphics[scale=0.43]{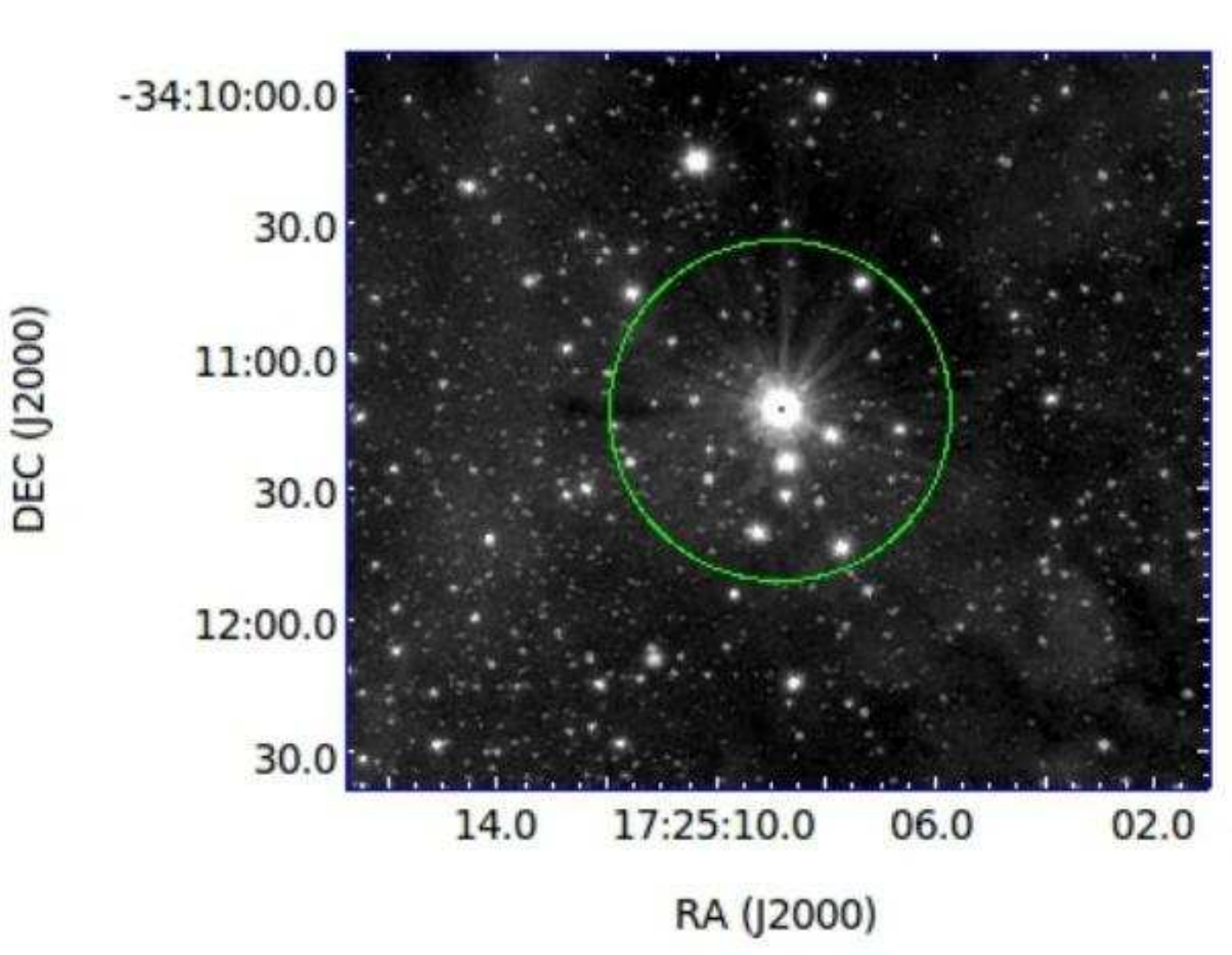}}
   \caption{Combined JHK$_S$ VVV images of clusters in the nebula NGC\,6357. Top panels: Pismis\,24 (left), BDS\,101 (right).
   Bottom panels: newly found clusters VVV\,CL164 (left) and VVV\,CL167 (right). The circles indicate 
   the angular sizes of the clusters (Table~\ref{table1}).}
   \label{Fig_VVV}
  \end{figure*}
 
 Several studies on NGC\,6357 are available in the literature, in particular on the cluster Pismis\,24.
 The WEBDA\footnote{http://www.univie.ac.at/webda/webda.html} database locates the center of Pismis\,24 at $\alpha$(2000)=17$^h$25$^m$32$^s$ and $\delta$(2000)=$-34$\textdegree25$'$00$''$, and provides a distance
 from the Sun d$_\odot\sim$1.99 kpc, reddening E(B-V)=1.72, and age of 10 Myr. WR\,93 has been considered a member of Pismis\,24 \citep{Massey_2001}. However, this star is located 4$'$ away from the cluster center. We investigate further this issue.
 
 To the southeast ($\alpha$(2000)=17$^h$25$^m$32$^s$ and $\delta$(2000)= $-34$\textdegree24$'$20$''$) of the complex is located an object first reported as ESO392-SC11 \citep{1982euse.book.....L}. It has been as well reported as AH03 J1725-34.4 \citep{2003stcl.book.....A}, BDS 100 \citep{2003A&A...404..223B}, G351.1+0.7 cluster \citep{2006RMxAC..26..180D}, and is also indicated in \citet{2011MNRAS.411..705M}. This extended ($\sim$3.5$'$) overdensity appears to be related to 
 the general young population of the complex. It is classified as a star cluster, see e.g. the DAML02 database\footnote{http://www.astro.iag.usp.br/ocdb/file/clusters.txt} \citep{2002A&A...389..871D}. They employed the designation AH03 J1725-34.4 and show an apparent diameter of 2.6$'$, a distance of 290 pc (too short for NGC 6357 complex) and an age of 7 Myr. In the present paper we use VVV data to constrain not only parameters, but the nature itself of this object as a young stellar cluster or an OB association (Sect.~\ref{ESO392}). In the present study we also focus our attention on the compact cluster BDS\,101 (Table~\ref{table1}) found by \citet{2003A&A...404..223B}, which is located at $\sim$1$'$ NW of the center of ESO392-SC11. 
 
 We show in Table~\ref{table1} the star clusters analysed in this work. Four star clusters were discovered by two of us (E. L. and E. B.). They are given VVV cluster identifications, in continuation to the recent cluster numbering by \citet{borissova}. They span the interval VVV\,CL164 to VVV\,CL167, according to the VVV designations.
 
 In Fig.~\ref{global} we show the angular distribution of the studied objects on a DSS image of the nebular complex NGC 6357. We can identify the presence of the stellar clusters in shell-like features from \citet{Cappa_2011}. Pismis\,24 (together with VVV\,CL165 and VVV\,CL166) is located near the border of G353.12+0.86. VVV\,CL167 might be related to a ring nebula. Finally, BDS\,101 and ESO\,392-SC\,11 are located in G353.1+0.6.
 
 In Fig.~\ref{Fig_VVV} we show a VVV $JHK_s$ blow up image of three embedded clusters and one open cluster in the sample. We emphasize the difference between embedded and open stellar clusters, the former are embedded in parental molecular clouds and the stellar content is in general dominated by PMS stars \citep[e.g.][]{2003ARA&A..41...57L, 2006A&A...445..567B}. On the other hand, OCs are dynamical survivours of the embedded phase \citep{2006MNRAS.369L...9B}. Pismis\,24 has very massive stars in the central region and attached to the North is the bright dust/HII region G353.2+0.9 \citep{Cappa_2011}. BDS101 is a compact EC while the loose one VVV\,CL167 is older. VVV\,CL164 is an OC projected in the NGC\,6357 area (Sect.~\ref{LBB2}).
    
 \begin{table*}
 \caption{Star clusters towards NGC\,6357.} \label{table1}
 \centering
 \begin{tabular}{lccccccc}
 \hline \hline
 Cluster&$\alpha$&$\delta$&l&b&R&Comments\\
 &(J200)&(J200)&(\textdegree)&(\textdegree)&(arcmin)& \\
 (1)&(2)&(3)&(4)&(5)&(6)&(7)\\
 \hline
 Pismis\,24\tablefootmark{a} &17:24:44 & -34:11:56 & 353.17  & 0.89 & 1.5 & optical EC\\
 BDS\,101\tablefootmark{b}  &17:25:34 & -34:23:09 & 353.11  & 0.65 & 0.7 & infrared EC \\
 ESO\,392-SC\,11\tablefootmark{c}&17:25:32& -34:24:20& 353.09& 0.64& 1.7 & EC or OB association?\\
 VVV\,CL164\tablefootmark{d}   &17:24:31 & -34:13:15 & 353.12 & 0.92  & 0.5  & infrared OC\\
 %Lima 1\tablefootmark{c} &17:24:49.0 & $-$34:14:32.0 & 353.1427 & 0.8552 & 0.15 & \\infrared EC
 VVV\,CL165\tablefootmark{d} &17:24:45 & -34:11:28 &  353.18 & 0.89 & 0.2 & subcluster of Pismis\,24?\\
 VVV\,CL166\tablefootmark{d} &17:24:47 & -34:12:36 &  353.17 & 0.88 & 0.2 & subcluster of Pismis\,24? \\
 VVV\,CL167\tablefootmark{d}&17:25:09 & -34:11:13 &  353.23 & 0.83 & 0.6  & infrared EC surrounding WR 93\\
 \hline
 \end{tabular}
 \tablefoot{Cols. 2 to 5: Optimised central coordinates. Col. 6: The radii were set by eye for decontamination purposes (Sect.~\ref{FSD}).
 \tablefoottext{a}{\citet{Moffat_1973}}
 \tablefoottext{b}{\citet{2003A&A...404..223B}}.
 \tablefoottext{c}{Sect. \ref{sc_6357}
 \tablefoottext{d}{Discovered in the present paper}.
 }
  }
 \end{table*}

 %%%%%%%%%%%%%%%%%%%%%%%%%%%%%%%%%%%%%%%%%%%%%%%%%%%%%%%%%%%%%%%%%%%%%%%%%%%%%%%%%%%%%
 
 \section{VVV photometry and data}
 \label{VVV_data}

  VVV is an ESO Public Survey
  scanning the Galactic bulge and plane with the 4-m class VISTA Telescope at Paranal
  \citep{Minniti_2010}. The total observed area is about 562 deg$^2$, scanning
  -10.0\textdegree $\lesssim b \lesssim$ +10.5\textdegree\ 
  and -10.3\textdegree $\lesssim l \lesssim$ +5.1\textdegree\
  in the bulge, and within 294.7\textdegree $\lesssim l \lesssim$ 350.0\textdegree\ and 
  -2.25\textdegree$\lesssim b\lesssim$ +2.25\textdegree\ in the plane. 
  The VVV survey observes in five passbands, namely $Z$ (0.87 $\mu$m), $Y$
  (1.02 $\mu$m), $J$ (1.25 $\mu$m), $H$ (1.64 $\mu$m), and $K_S$ (2.14 $\mu$m).
  It also conducts a variability campaign in the $K_S$-band only, with
  $\sim$100 pointings spanning six years (2010–2016). The JHK$_S$ observations were
  completed during the first semester of 2011. The present work is based on J, H
  and K$_S$ photometry, from the VVV Data Release 1 (DR1) \citep{Saito_2012}.

  Each unit of VISTA observations is called a “tile”,
  consisting of six individual pointings (or ``pawprints'')
  and covers a 1.64 deg$^2$ field of view. To fill up the VVV area, a
  total of 348 tiles are used, with 196 covering the bulge (a
  $14\times14$ grid) and 152 the Galactic plane (a $4\times38$ grid). 
  The NGC\,6357 complex appears in the VVV bulge/disk tiles b329 and b343. 
  
  Photometric catalogues for the VVV images are provided by
  the Cambridge Astronomical Survey Unit (CASU)\footnote{http://casu.ast.cam.ac.uk/vistasp/}.
  The catalogues contain the positions, magnitudes, and some shape measurements 
  obtained from different apertures, with a flag indicating
  the most probable morphological classification. The limiting magnitude
  for the aperture photometry of the catalogues occurs at
  K$_S$=18 mag in most disk fields. The VVV data are in the natural 
  VISTA Vegamag system, with the photometric calibrations in $JHK_S$ performed 
  using the VISTA magnitudes of unsaturated 2MASS\footnote{The Two Micron All Sky Survey, 
  All Sky data release \citep{Skrutskie_2006}} stars present in the
  images. The present work is based on the derived colours and magnitudes 
  established by the CASU v1.3 pipeline reduction. The detailed account of the CASU pipeline
  can be found in \citet{Irwin_2004}.
  
  Only sources with VVV K$_S$ photometry defined as “stellar” (sources with a Gaussian 
  sigma parameter between 0.9 and 2.2) were selected. Saturated stars in the VVV data, usually brighter
  than J, H and K$_S$ $\sim$11 mag, were replaced by the respective 2MASS magnitudes.

 %%%%%%%%%%%%%%%%%%%%%%%%%%%%%%%%%%%%%%%%%%%%%%%%%%%%%%%%%%%%%%%%%%%%%%%%
 
 \section{Star cluster analyses}
 \label{SCA}
 
 Besides analysing open clusters \citep[e.g.][]{2009MNRAS.392..483B}, our group has also concentred efforts 
 to study in detail embedded clusters, by developing tools to extract information from the CMDs and RDPs 
 \citep[e.g.][]{2009MNRAS.397.1915B, 2010A&A...516A..81B}.
 
 NGC\,6357 in itself is an astrophysical laboratory, not only for its numerous interstellar structures, e. g.
 filaments, bubbles, knots, etc, but also for its ECs (Table~\ref{table1}). VVV photometry provides an adequate means to explore 
 them, in such a crowded fields towards the central disk and bulge.
 
 \subsection{Field star decontamination}
 \label{FSD}

  Field-star decontamination is usually required for the 
  identification and characterisation of star clusters.
  To disentangle field and cluster stars we use a statistical
  decontamination algorithm \citep{2007A&A...473..445B, 2010A&A...516A..81B} adapted to the 
  photometric depth of VVV.
  The comparison fields used for the decontamination depend on
  the projected distribution of individual stars and the presence of other clusters and/or clumpy
  extinction due to dark clouds in the area. Examples are a ring around the cluster or
  some other comparison field selected in its vicinity.
  The algorithm measures the relative number densities 
  of probable field and cluster stars in cubic CMD
  cells with axes along the J magnitude and
  (J-H) and (J-K$_S$) colours. It (i) divides the 
  range of CMD magnitude and colours into a 3D grid, 
  (ii) estimates the number density of field stars in each cell based
  on the number of comparison field stars with similar magnitude and
  colours as those in the cell, and (iii) subtracts the expected number
  of field stars from each cell. Input algorithm parameters are the cell
  dimensions $\Delta$J=1.0 and $\Delta$(J-H)=$\Delta$(J-K$_S$)=0.2 mag.
  Summing over all cells, each grid setup produces a total
  number of member stars $\langle$N$_{mem}\rangle$ and, repeating this
  procedure for the 729 different setups (different cell sizes and their positionings), we obtain the average number of members $\langle$N$_{mem}\rangle$. Each star is ranked according
  to the number of times they survive all runs (survival frequency) and
  only the $\langle$N$_{mem}\rangle$ highest ranked stars are accepted as cluster
  members and transposed to the respective decontaminated CMD.

  We decontaminated CMDs to investigate the nature
  of star cluster candidates and derive their astrophysical parameters.
  In summary, we applied (i) field-star decontamination to
  uncover the intrinsic CMD morphology, essential for derivation of reddening, age, and distance to the Sun, and (ii)
  colour-magnitude (CM) filters (Fig.~\ref{cmd_pismis24}) to exclude stars unlike those of the CMD sequence.
  The latter filters are wide enough to include cluster MS and PMS stars, together with the photometric uncertainties and binary star effects (Sect.~\ref{CMDs}). In the following sections we will make use of this tool several times in the analyses.
  The latter procedure is required for intrinsic stellar RDPs. In  particular, the use of field-star
  decontamination in the construction of CMDs has proved to constrain age and distance
  much more than the raw (observed) photometry \citep[e.g.][]{2010A&A...516A..81B}.

%%%%%%%%%%%%%%%%%%%%%%%%%%%%%%%%%%%%%%%%%%%%%%%%%%%%%%%%%%%%%%%%%%%%% 
  \subsection{Colour-Magnitude Diagrams}
  \label{CMDs}
  
  CMDs built with the raw photometry of the present objects are shown in the top panels of Figs.~\ref{cmd_pismis24},~\ref{cmd_bbd95},~\ref{cmd_lbb2} and ~\ref{cmd_ESO}. For qualitative comparison, CMDs extracted from equal-area 
  comparison fields are shown in the middle panels. The decontamination itself is based on as large as possible field areas. Finally, the decontaminated CMDs are shown in the bottom panels, together with the respective CM filters. 
  
  Fundamental parameters (Table \ref{tb_par}) are derived by means of the constraints provided by the field-decontaminated 
  CMD morphologies combining the MS and PMS distributions. Historically, different approaches have been
  used to extract astrophysical parameters from isochrone fits. A review of these methods is given
  by \citet{2006MNRAS.373.1251N}. In the present cases fits are matched \textit{by eye}, taking the combined
  MS and PMS stellar distribution as constraint. Throughout the paper we use Padova isochrones with solar metallicity (Z=0.0019) \footnote{http://stev.oapd.inaf.it/cgi-bin/cmd} 
  \citep{2012MNRAS.427..127B} computed for the VISTA Z, Y, J, H  and K$_S$ filters. 
  The derived fundamental parameters are given in Table \ref{tb_par}, where we also provide the Galactocentric 
  distance (R$_{GC}$), which is based on the derived value of the Sun's distance to the 
  Galactic center R$_{\odot}$=7.2 kpc, computed by means of Globular clusters \citep{2006A&A...450..105B}.
  We obtained d$_{\odot}$=1.78$\pm$0.1 kpc for the NGC\,6357 complex based on the individual determinations for Pismis\,24, BDS\,101, ESO\,392-SC\,11 and VVV\,CL167 (Table~\ref{tb_par}).

\begin{table*}
 \caption{Derived fundamental parameters for the star clusters towards NGC\,6357.} \label{tb_par}
 \centering
 \begin{tabular}{lccccccccc}
 \hline \hline
 Cluster&A$_V$&Age&d$_{\odot}$&R$_{GC}$\\
 &(mag)&(Myr)&(kpc)&(kpc)\\
 (1)&(2)&(3)&(4)&(5)\\
 \hline
 Pismis\,24 & 5.87$\pm$0.06 & 5$\pm$2 & 2.0$\pm$0.1 & 5.3$\pm$0.1\\
 BDS\,101   & 6.57$\pm$0.06 & 5$\pm$2 & 1.7$\pm$0.1 & 5.6$\pm$0.1\\
 ESO\,392-SC\,11& 6.69$\pm$0.06 &5$\pm$2 & 1.9$\pm$0.1 & 5.4$\pm$0.1\\
 VVV\,CL164 & 15.9$\pm$0.09 & $(5\pm2)\times10^3$ & 1.4$\pm$0.1&5.9$\pm$0.1\\
 %VVV\,CL165 &  &  &\\
 %VVV\,CL166 &  &  &\\
 VVV\,CL167 & 4.94$\pm$0.06 & 9$\pm$2 & 1.6$\pm$0.1& 5.6$\pm$0.1\\
 \hline
 \end{tabular}
 \tablefoot{Table Notes. Col. 2: $A_V$ absorption in the cluster central region. Col. 3: age, from VVV
photometry. Col. 4: distance to the Sun. Col. 5: R$_{GC}$ distance of the object to the Galactic center.}
\end{table*} 

 \subsubsection{Pismis\,24}
   
\begin{figure}
 \centering
{\includegraphics[scale=0.6, clip=true]{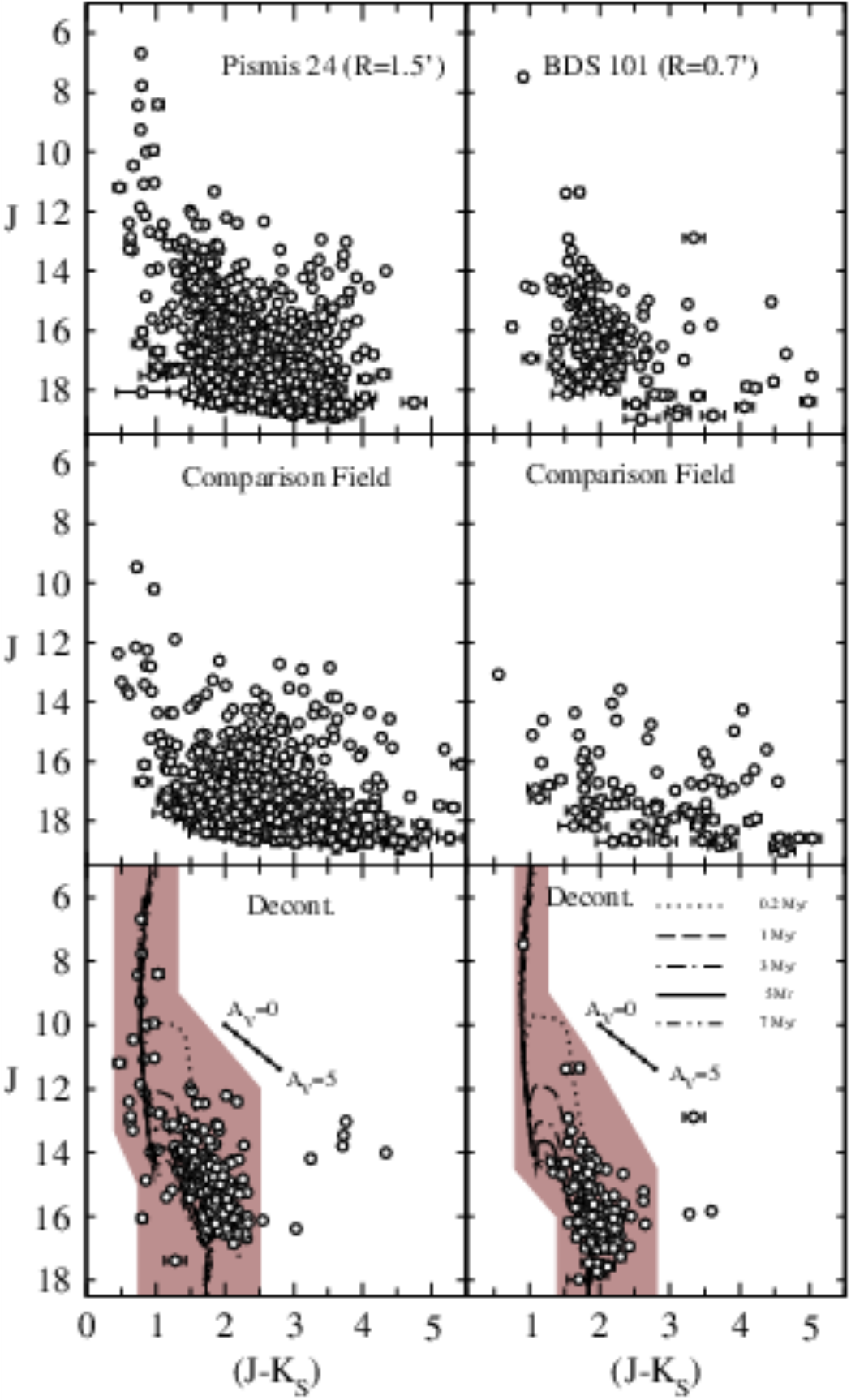}}
 \caption{Top left panel: Jx(J-K$_S$) CMDs of Pismis\,24 showing the observed 
 photometry for representative cluster and field regions. Middle left panel: equal-area extraction
 from the comparison field. Bottom left panel: The decontaminated CMDs fitted with Padova isochrones 0.2, 1, 3, 5 and 7 Myr, together with the Colour-Magnitude filter (brown polygon) used to isolate the MS and PMS stars. Right panels: the same for BDS\,101. Reddening vector for A$_v$=0$-$5 mag is shown.}
 \label{cmd_pismis24}
 \end{figure}

 The decontaminated CMD of Pismis\,24 (Fig.~\ref{cmd_pismis24}) presents a relatively vertical
 populous MS. It has a large population of red faint stars belonging to the PMS, making it a rather 
 massive object in the sample. This cluster contains 12 known massive stars,
 three of them (Pis24-1NE, Pis24-1SW and Pismis\,24-17) with $\sim$100 M$_\odot$ each
 \citep{2007ApJ...660.1480M}. These stars are absent in 2MASS and are saturated in
 VVV. Consequently, they do not appear in our CMDs.
 There is a debate in the literature on the distance of Pismis\,24, and published estimates 
 range from 1.0 to 3.0 kpc.
 We fitted a set of Padova isochrones (0.2, 1, 3 and 5 Myr) (Fig. \ref{cmd_pismis24}) leading
 to $E(J-K_S)=1.01\pm0.01$ ($E(B-V)=1.75\pm0.10$ or 
 $A_V=5.87\pm0.06$). The observed and absolute
 distance moduli are $(m-M)_J=13.20\pm0.10$ and $(m-M)_O=11.50\pm0.10$, respectively, and 
 $d_{\odot}=2.0\pm0.1$ kpc. We estimated the mass of the PMS population of Pismis\,24 by counting the
number of PMS stars and multiplying it by a mean PMS stellar mass. We obtained 153 M$_{\odot}$.
For the mean PMS stellar mass, we assumed an initial mass function of \citet{2001MNRAS.322..231K}
between 0.08 and 7 M$_{\odot}$, which results in a mean mass of 0.6 M$_{\odot}$ \citep{2010A&A...516A..81B}. For the MS stars in Fig.~\ref{cmd_pismis24} the sum of the stellar masses is 80 M$_{\odot}$. Finally, we added the three saturated supermassive stars not included in the VVV photometry, totalling 533$\pm$50 M$_{\odot}$. We conclude that Pismis\,24 is not an extremely massive cluster, having an intermediate mass which is mostly stocked in its massive stars.
  
 \subsubsection{BDS\,101}
 
 BDS\,101 is a compact cluster. Differently from Pismis\,24 its CMD (Fig 
 \ref{cmd_pismis24}) shows a poorly-populated MS and a rich PMS. The single star that occupies the MS
 is an O5\footnote{SIMBAD} star. We fitted a set of Padova isochrones (0.2, 1, 3, 5 and 7 Myr) (Fig. \ref{cmd_pismis24}) 
 leading to $E(J-K_S)=1.13\pm0.01$ ($E(B-V)=1.95\pm0.10$ or 
 $A_V=6.57\pm0.06$). The observed and absolute
 distance moduli are $(m-M)_J=13.0\pm0.10$ and $(m-M)_O=11.09\pm0.10$, respectively, and 
 $d_{\odot}=1.66\pm0.09$ kpc. The computed parameters are consistent with those of Pismis\,24, indicating that BDS\,101 is also embedded
 in the NGC\,6357 complex. The total mass estimate for BDS\,101 is 106$\pm$10 M$_{\odot}$. Although a prominent cluster, it is not massive.

 \subsubsection{VVV\,CL167}
 
 \begin{figure}
 \centering
 {\includegraphics[scale=0.6, clip=true]{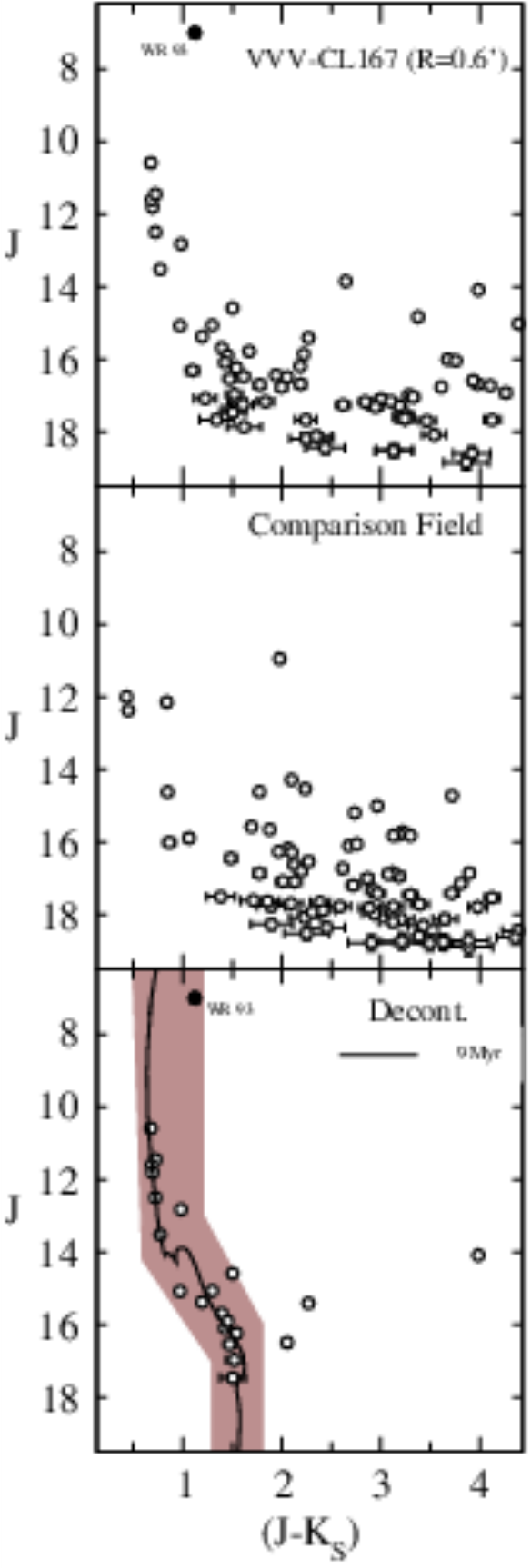}}
 \caption{Top panel: observed Jx(J-K$_S$) CMD of the region R<0.6$'$ of VVV\,CL167. 
 Middle left panel: equal-area extraction from the comparison field. Bottom panel: 
 the decontaminated CMD best-fit Padova isochrone is 9 Myr (solid line) 
 together with the CM filter (brown polygon). WR\,93 is indicated.}
 \label{cmd_bbd95}
 \end{figure} 
 
 \begin{figure}
\centering
{\includegraphics[scale=0.5, clip=true]{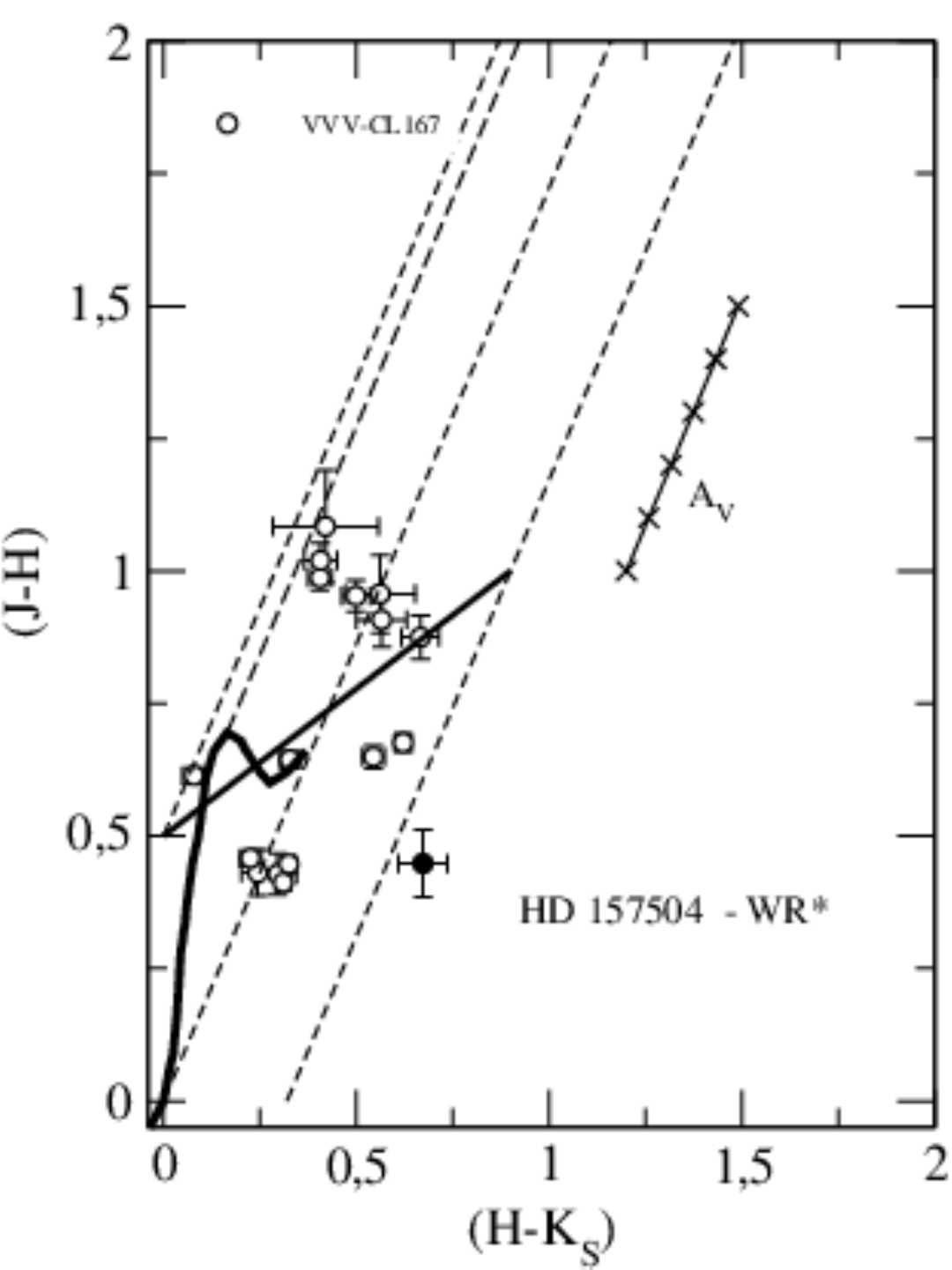}}
\caption{Colour-colour diagram ((J-H)x(H-K$_S$)) of the decontaminated stars in VVV\,CL167. The continuous line 
represents the intrinsic distribution of spectral types and the continuous straight line indicates the undereddened
locus of T Tauri stars. The redenning vector corresponds to A$_V$=5. Reddening directions for M5 giants, O3 and T Tauri stars are
shown as dashed lines.}
\label{dcc_bbd95}
\end{figure} 
 
 We find that this cluster contains the WR star WR\,93 \citep{Van_der_Hucht_01}. 
 Historically, this star has been considered a likely member of Pismis-24 \citep{Massey_2001}. WR stars usually have masses in the range 10-25 $M_{\odot}$, and present strong, broad emission lines of He and N (WN) or He, C
 and O (WC) \citep{2005A&A...429..581M}.
 Recently, clusters containing WR stars were studied with VVV \citep{2012A&A...545A..54C}. We point out that WR\,93 is located too far ($\sim$5$'$ West) from Pismis\,24. 
 %At its distance, 5 arcmin corresponds to 4 pc, while stellar drifts at 10 km.s$^{-1}$ over 1 Myr would reach to 10 pc.
 The image (Fig.~\ref{Fig_VVV}) and the CMD (Fig.~\ref{cmd_bbd95}) indicate a star cluster surrounding WR 93. We identify very few 
 PMS stars in the CMD (Fig. \ref{cmd_bbd95}). We adopted as best fit a 
 Padova isochrone of 9 Myr (Fig. \ref{cmd_bbd95}) that produces
 $E(J-K_S)=0.85\pm0.01$ ($E(B-V)=1.47\pm0.10$ or $A_V=4.94\pm0.06$). We get the observed and absolute
 distance moduli $(m-M)_J=12.50\pm0.10$ and $(m-M)_O=11.07\pm0.10$, respectively. The distance to the Sun is 
 $d_{\odot}=1.6\pm0.1$ kpc. 
 In the colour-colour diagram (Fig.~\ref{dcc_bbd95}) the star WR\,93 shows K$_S$-excess.  A similar K$_S$-excess was
  observed for massive stars in the $\sim14$\,Myr open cluster NGC 4755 \citep{2006A&A...453..121B}.  
 The CMD of VVV\,CL-167 contains essentially only MS stars and WR\,93. We obtain a cluster mass of 50$\pm$5 M$_{\odot}$. The cluster age and mass point to a dynamically evolved cluster.
 
 \subsubsection{VVV\,CL164 -  a projected OC next to the complex} 
 \label{LBB2}

\begin{figure}
 \centering
 {\includegraphics[scale=0.73, clip=true]{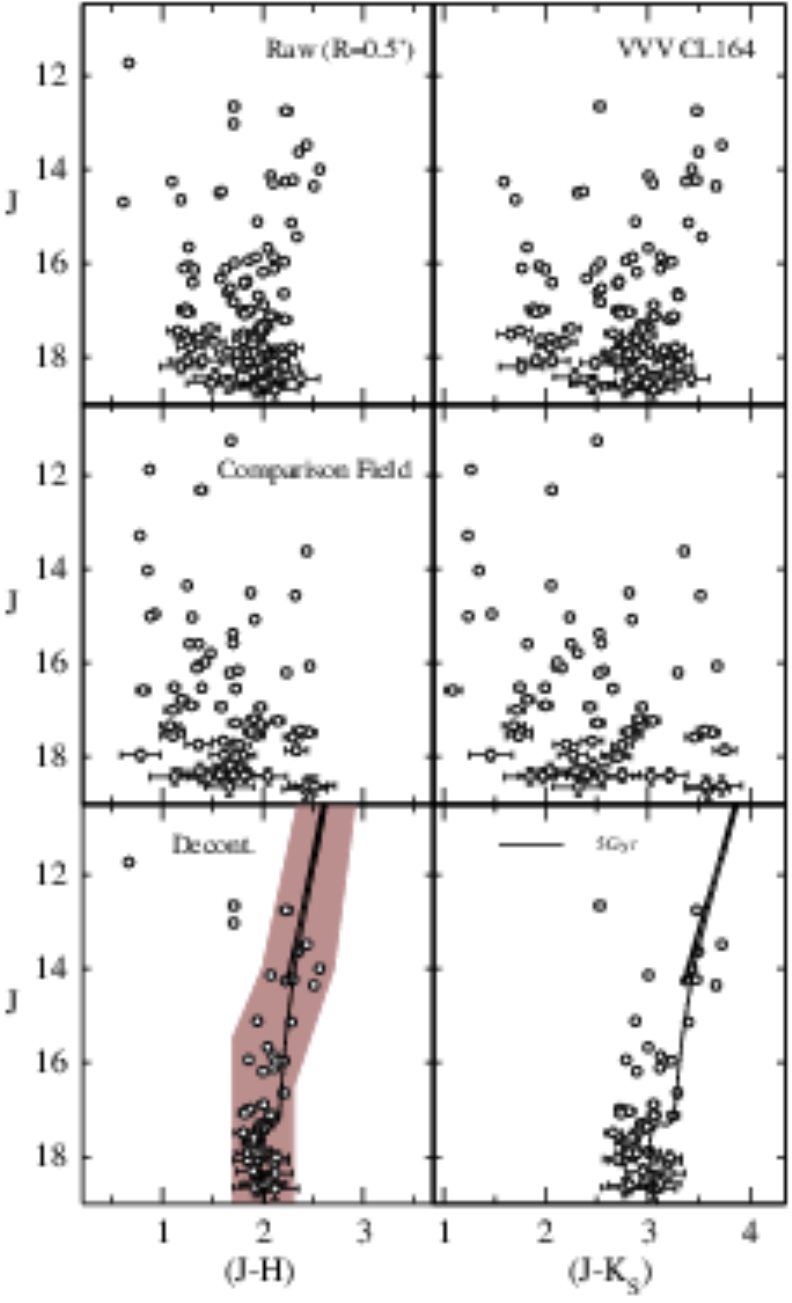}}
 \caption{Top left panel: observed Jx(J-H) CMD of the region R<0.5 arcmin of VVV\,CL164. 
 Middle left panel: equal-area extraction from the comparison field. Bottom left panel: 
 the decontaminated CMD best-fit Padova isochrone (5 Gyr) together with the CM filter (brown polygon) used to isolate the MS and giants. Right panels: the same for Jx(J-K$_S$).}
 \label{cmd_lbb2}
 \end{figure}

 The decontaminated CMDs Jx(J-H)and Jx(J-K$_S$) are shown in the bottom panels of Fig.~\ref{cmd_lbb2}. We also include the CM filter on the Jx(J-H) decontaminated CMD. The remaining stars populate sequences of typical intermediate-age open clusters, with evidence of a red clump and a turn-off (TO) in both colours.
 Blue stragglers appear to be present too. VVV\,CL164 presents relatively large errors at the TO level (J $\sim$18--19 mag).
 We matched a set of Padova isochrones (ages between 3 and 7 Gyr) and adopted 5\,Gyr as the best solution (Fig. \ref{cmd_lbb2}). We obtain $E(J-H)=1.70\pm0.01$. ($E(B-V)=4.73\pm0.15$ or $A_V=15.9\pm0.1$). The A$_V$ value is considerably larger than for the other clusters. The observed and absolute
 distance moduli are $(m-M)_J=15.30\pm0.10$ and $(m-M)_O=10.69\pm0.10$, respectively, resulting
 $d_{\odot}=1.37\pm0.07$ kpc. We conclude that VVV\,CL164 is an intermediate-age cluster (5$\pm$2 Gyr) that is projected next to the NGC\,6357 complex.
 
\begin{figure}
\centering
{\includegraphics[scale=0.55, clip=true]{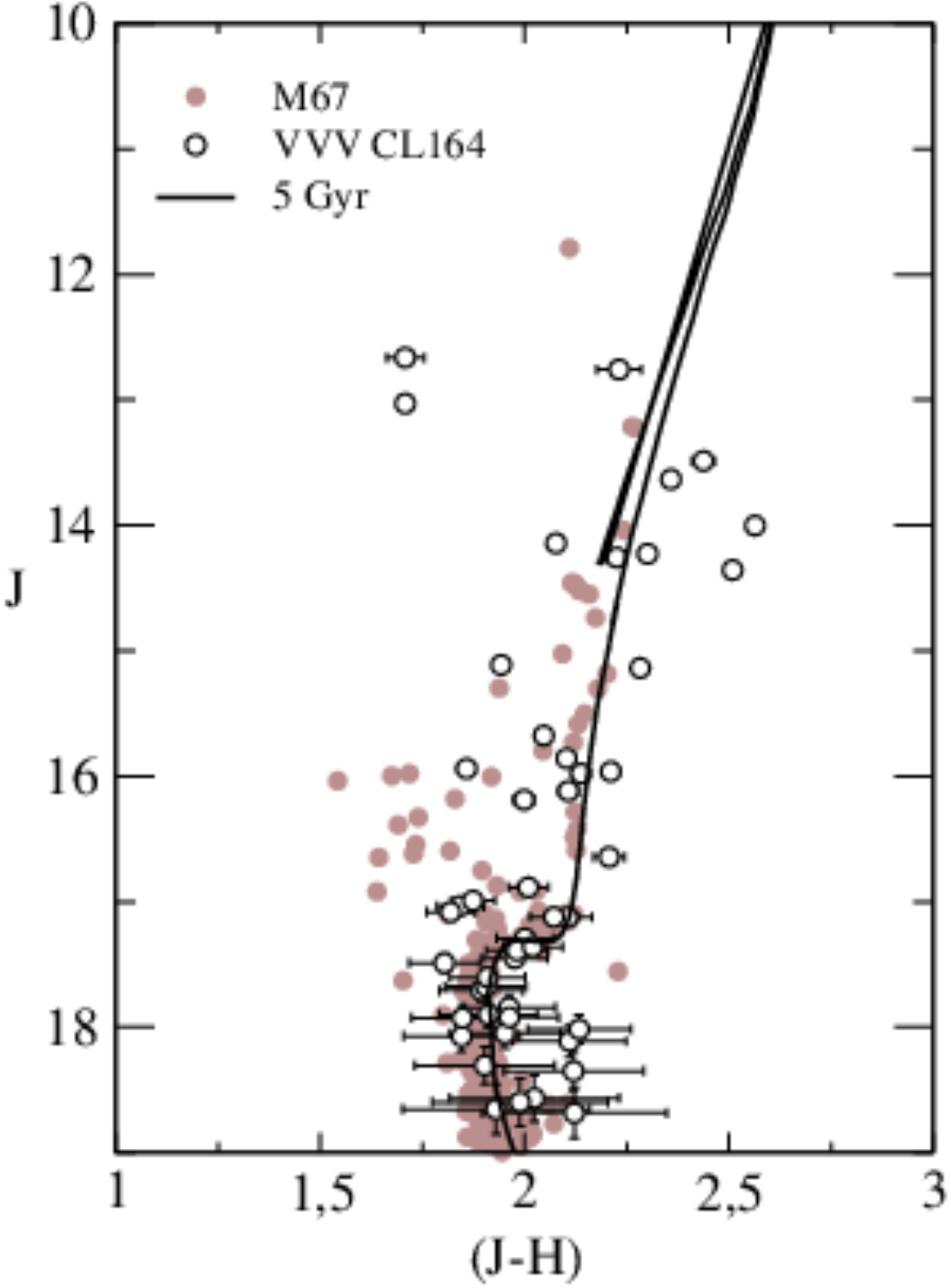}}
\caption{Open cluster VVV\,CL164: comparison between the decontaminated Jx(J-H) CMD for the region
R<0.5 arcmin of VVV\,CL164 (empty circles) and the region R<10 acrmin of M67 (filled circles). The best-fit to VVV\,CL164 is a Padova isochrone of 5 Gyr.}
\label{cmd_m67lbb2}
\end{figure}

 Contrasting with the young clusters in the complex, the CMD of VVV\,CL164 presents MS, TO
 and blue straggler stars, similar to those of M67 built with 2MASS data (Fig.~\ref{cmd_m67lbb2}). We estimated the mass of this relatively old open cluster by adding stars throughout the decontaminated CMD using the stellar masses for M\,67 according to \citet{2003A&A...405..525B}. We obtained $\approx$ 200 M$_{\odot}$ for VVV\,CL164.

\subsection{Additional case: subclustering in Pismis\,24?}
    
The spatial distribution of YSOs in a cluster provides
insights on the fragmentation processes leading to the formation
of protostellar cores, evidence for triggered star formation and
the subsequent dynamical evolution of the stars as they evolve from protostellar to the MS \citep{2004ApJS..154..367M}.    

 \begin{figure*}
\centering
{\includegraphics[width=4.8in, trim= .5 .5 .5 .5, clip=true]{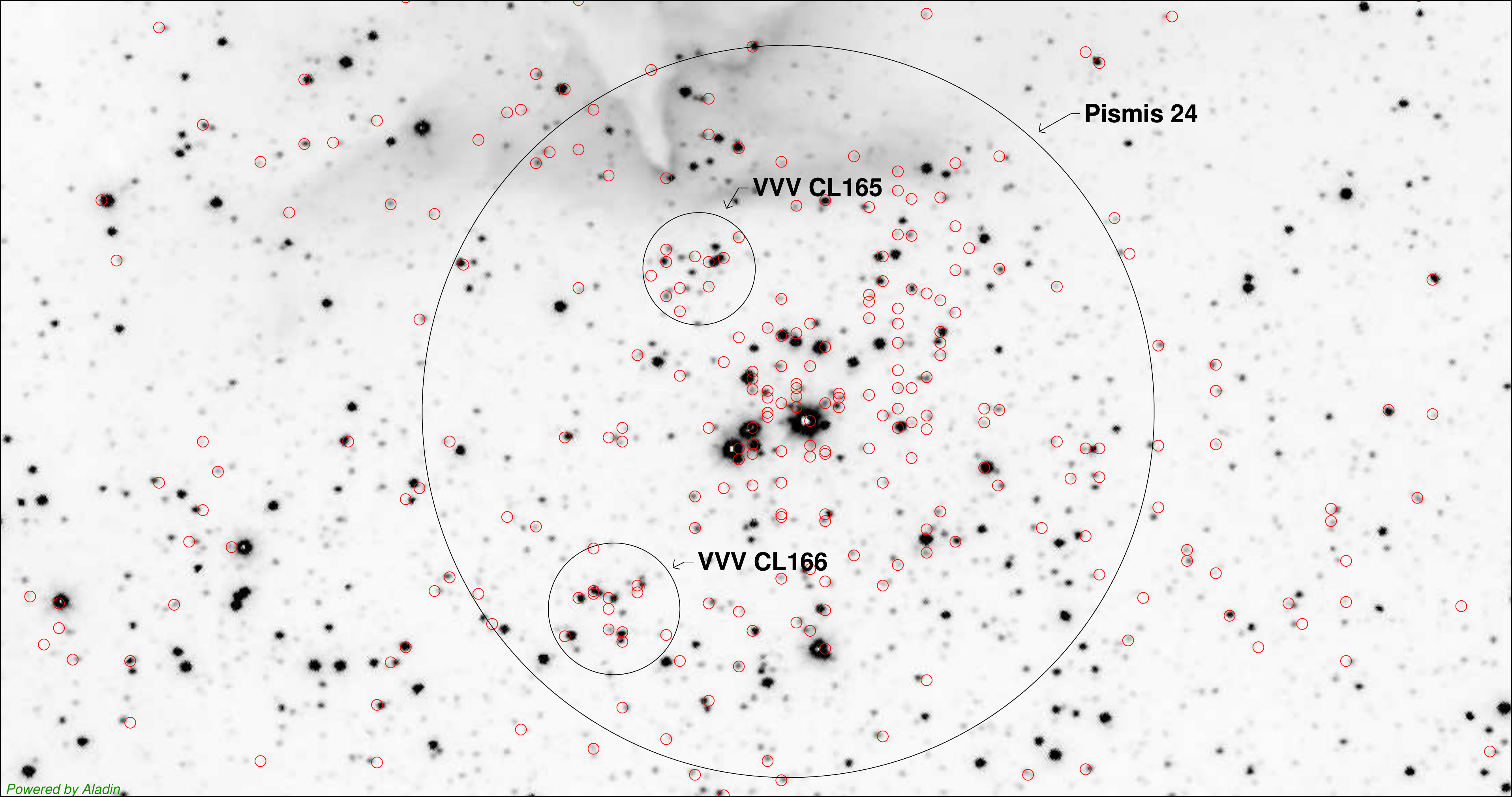}}
\caption{$5'\times3'$ VVV K$_S$ image of Pismis\,24 area (large circle). The probable subclusters VVV\,CL165 (top circle) and VVV\,CL166 (bottom circle) are indicated. 
The small red circles indicate YSOs \citep{Fang_2012}. The shift between the stars in the VVV image and the YSOs from Fang catalog (2MASS data) is a consequence of the difference of angular resolution of each survey, 0.34$''$ and 2.0$''$ respectively. The black circle radius is given in Table~\ref{table1}.}
\label{fang}
\end{figure*}

 \begin{figure*}
   \centering
   {\includegraphics[scale=0.5]{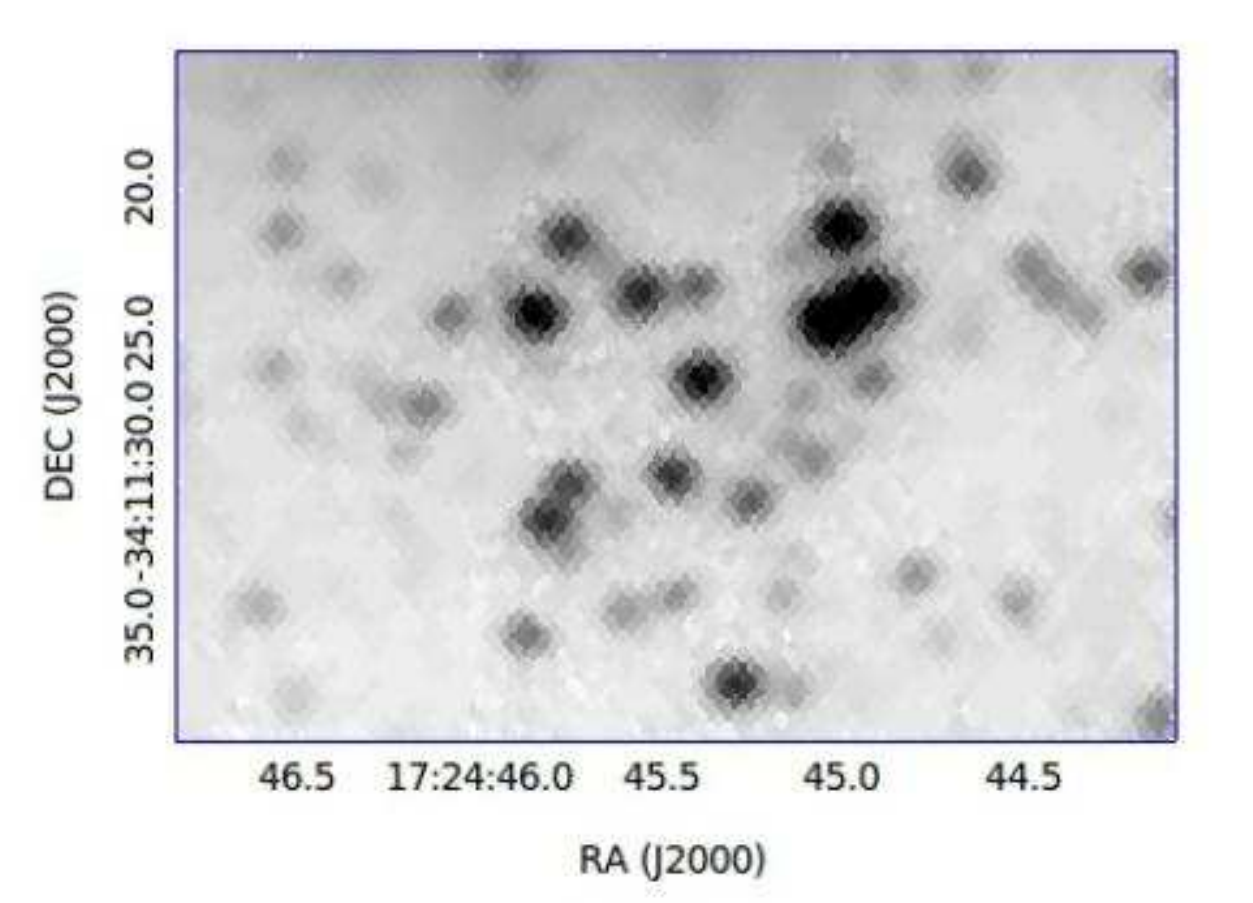}
   \includegraphics[scale=0.4]{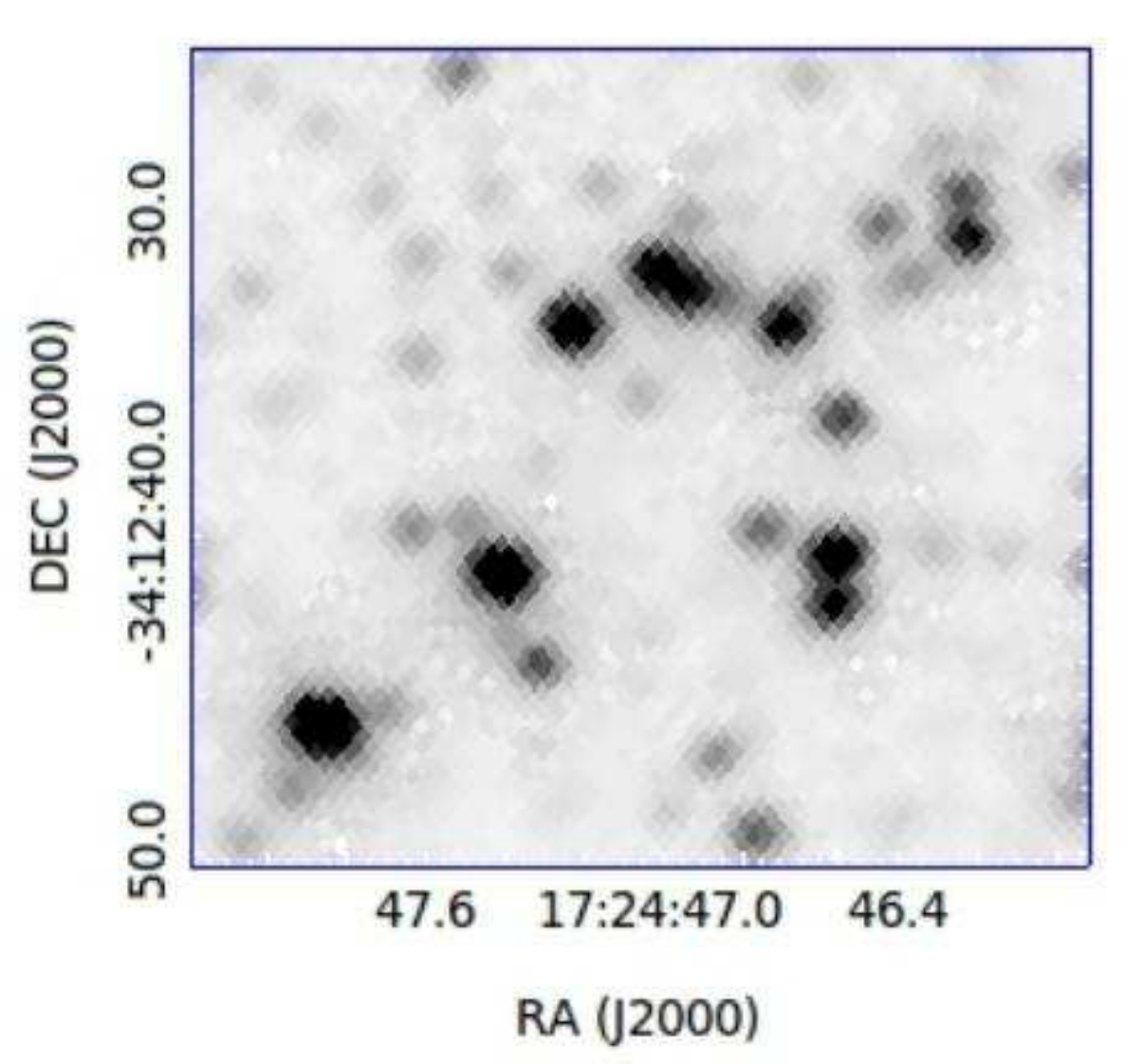}}
   \caption{K$_S$ VVV blow up images of the probable subclusters located in Pismis\,24. Left panel: VVV\,CL165. Right panel: VVV\,CL166.}
   \label{blowup_fangs}
  \end{figure*}

In Fig.~\ref{fang} we show the YSOs from \citet{Fang_2012} in Pismis\,24 area.
Pismis\,24 appears to have two close neighbours that are possible subclusters of YSOs. Blow up VVV images of these objects suggest clusters (Fig.~\ref{blowup_fangs}). The stellar density excesses in the RDP (Sect.~\ref{rdp_sect}) of Pismis\,24 produced by these two possible subclusters are indicated in 
Fig.~\ref{rdp}. The formation of these young stars may have been influenced by winds from the nearby 
OB stars in the core of Pismis\,24. A possible fate of the ensemble is merging with the more
massive cluster \citep{2007ApJ...655L..45M, 2013ApJ...764...73P}.

\subsection{The EC ESO\,392-SC\,11 (AH\,03 J1725-34.4)}
\label{ESO392}

\begin{figure}
\centering
{\includegraphics[width=3.0in, trim= .5 .5 .5 .5, clip=true]{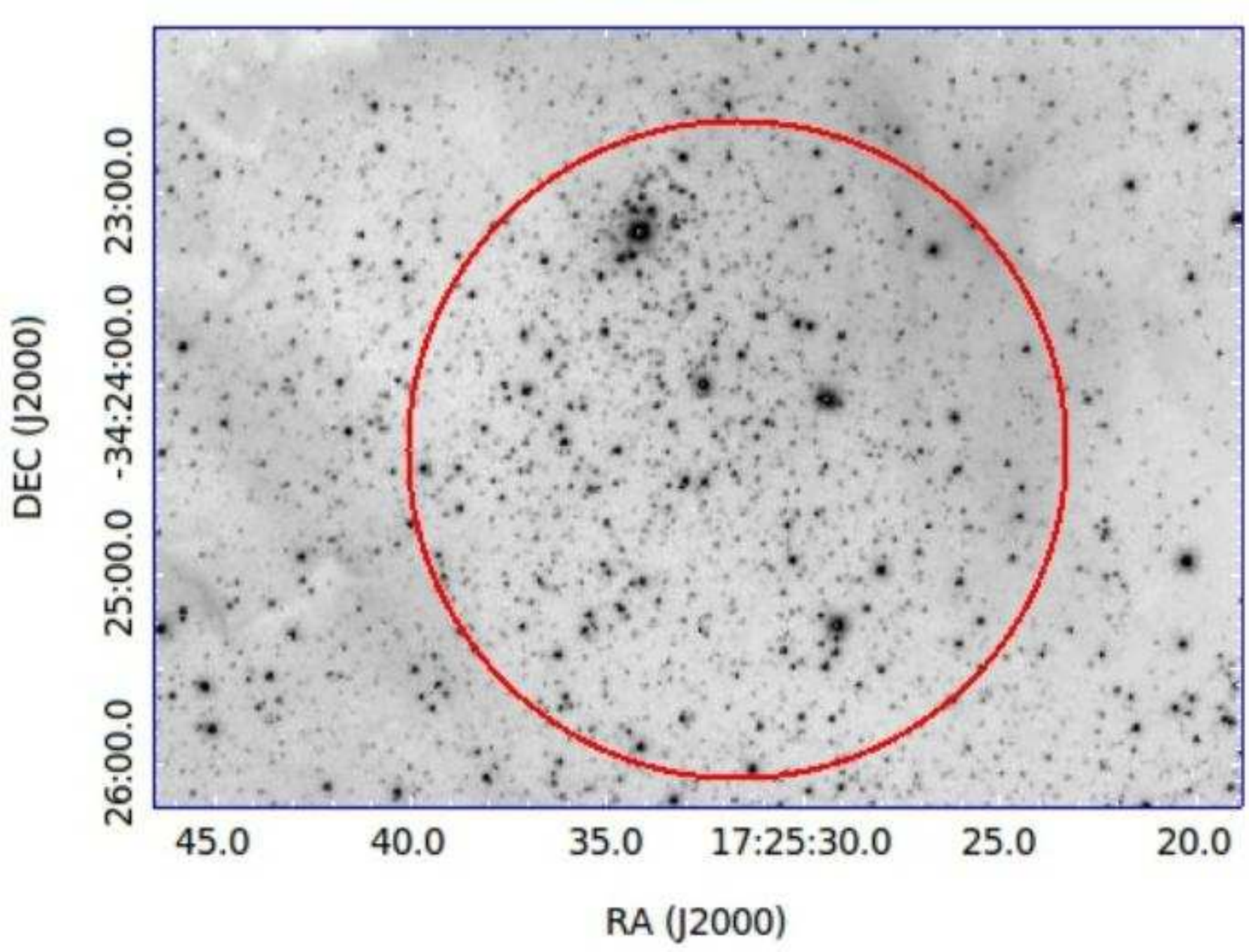}}
\caption{ K$_S$ VVV image of ESO392-SC11. The circle indicates the object area. The compact cluster near the northern edge is BDS\,101.}
\label{image_ESO}
\end{figure}

\citet{2011ApJ...728L..37R, 2011ApJ...743L..28R} have observed, in particular using GLIMPSE images and spectroscopy of luminous stars, far OB associations in the Galaxy which appear to be more massive than $10^4$ $M_{\odot}$. In Fig.~\ref{image_ESO} we show a VVV $K_s$ image of ESO\,392-SC\,11, where the compact cluster near the NW edge is BDS\,101. The RDP of ESO\,392-SC\,11 appears to be rather flat (Fig.~\ref{RDP_ESO}) like that of an association.

\begin{figure}
 \centering
 {\includegraphics[scale=0.7, clip=true]{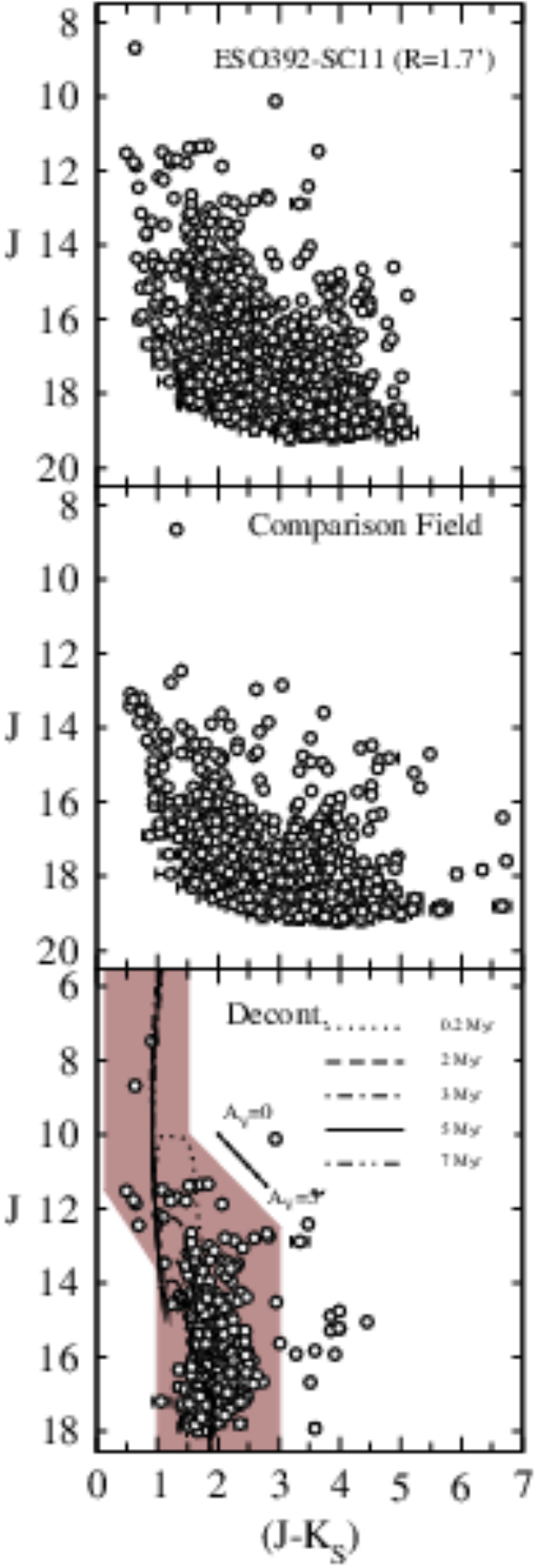}}
 \caption{Top panel: observed Jx(J-K$_S$) CMD of the region R<1.7 arcmin of ESO\,392-SC\,11. 
 Middle left panel: equal-area extraction from the comparison fields. Bottom panel: 
 the decontaminated CMD together with the CM filter (brown polygon). MS and PMS isochrones are given.}
 \label{cmd_ESO}
 \end{figure}

The decontaminated CMD Jx(J-K$_S$) is shown in the bottom panel of Fig.~\ref{cmd_ESO}. We fitted a set of Padova isochrones (0.2, 1, 3 and 5 Myr) leading to $E(J-K_S)=1.15\pm0.01$ ($E(B-V)=1.99\pm0.10$ or 
 $A_V=6.69\pm0.06$). The observed and absolute distance moduli are $(m-M)_J=13.3\pm0.10$ and $(m-M)_O=11.36\pm0.10$, respectively, resulting in  
 $d_{\odot}=1.87\pm0.10$ kpc. This set of parameters shows that ESO\,392-SC\,11 is part of NGC\,6357 complex, likewise Pismis\,24 and BDS\,101.
 
The off-Galaxy position of the Magellanic Clouds is suitable for size measurements of their stellar associations. \citet{1995ApJS..101...41B} and \citet{1999AJ....117..238B} measured size and position angle
for 3226 associations in the Magellanic system (LMC, SMC, and Bridge). In the LMC, \citet{1970AJ.....75..171L} reported diameters in the range 15-350 pc,
while in the SMC, \citet{1985PASP...97..530H} gave 20 to 140 pc. The smallest detected associations in the SMC \citep{1995ApJS..101...41B} and LMC \citep{1999AJ....117..238B} have diameters of 5 and 7 pc, respectively, while the largest ones have typically 200 pc. The RDP radius of ESO\,392-SC\,11 is 3.3$'$ (Table~\ref{tb_rdp}). This value corresponds to a diameter of 2.6\,pc, thus smaller than the lower limit of the associations in the Magellanic Clouds.

Some of the Magellanic Clouds associations were studied with HST reaching the PMS population. LH\,95 in the LMC has a diameter about 40 pc and includes a series of PMS clusters \citep{2007ApJ...665L..27G}. The total mass of LH 95 with HST is 3000 M$_{\odot}$ \citep{2012MNRAS.422.3356D}. ESO\,392-SC\,11 in diagnostic diagrams (Sect.~\ref{structure}) involving structural parameters and mass behaves like a star cluster, rather than an association. As a cautionary remark, the Magellanic Clouds and the Galaxy have different tidal environments that should affect fragmentation of the molecular clouds.
 
Since ESO392-SC11 is made up of PMS stars and contains only a few  MS stars,
its association appearence will certainly change to one of a star cluster
when more PMS stars evolve to the MS. Thus our results suggest that ESO392-SC11
is in a pre association stage. If the region loses gas and dust, then the system
may be destabilized, later dynamically evolving to an association, after a star cluster phase.

%%%%%%%%%%%%%%%%%%%%%%%%%%%%%%%%%%%%%%%%%%%%%%%%%%%%%%%%%%%%%%%%%%%%%%%%%%%%

 \subsection{Stellar density profiles}
 \label{rdp_sect}
 
\begin{figure}
  \centering
  {\includegraphics[width=3in, angle=-90]{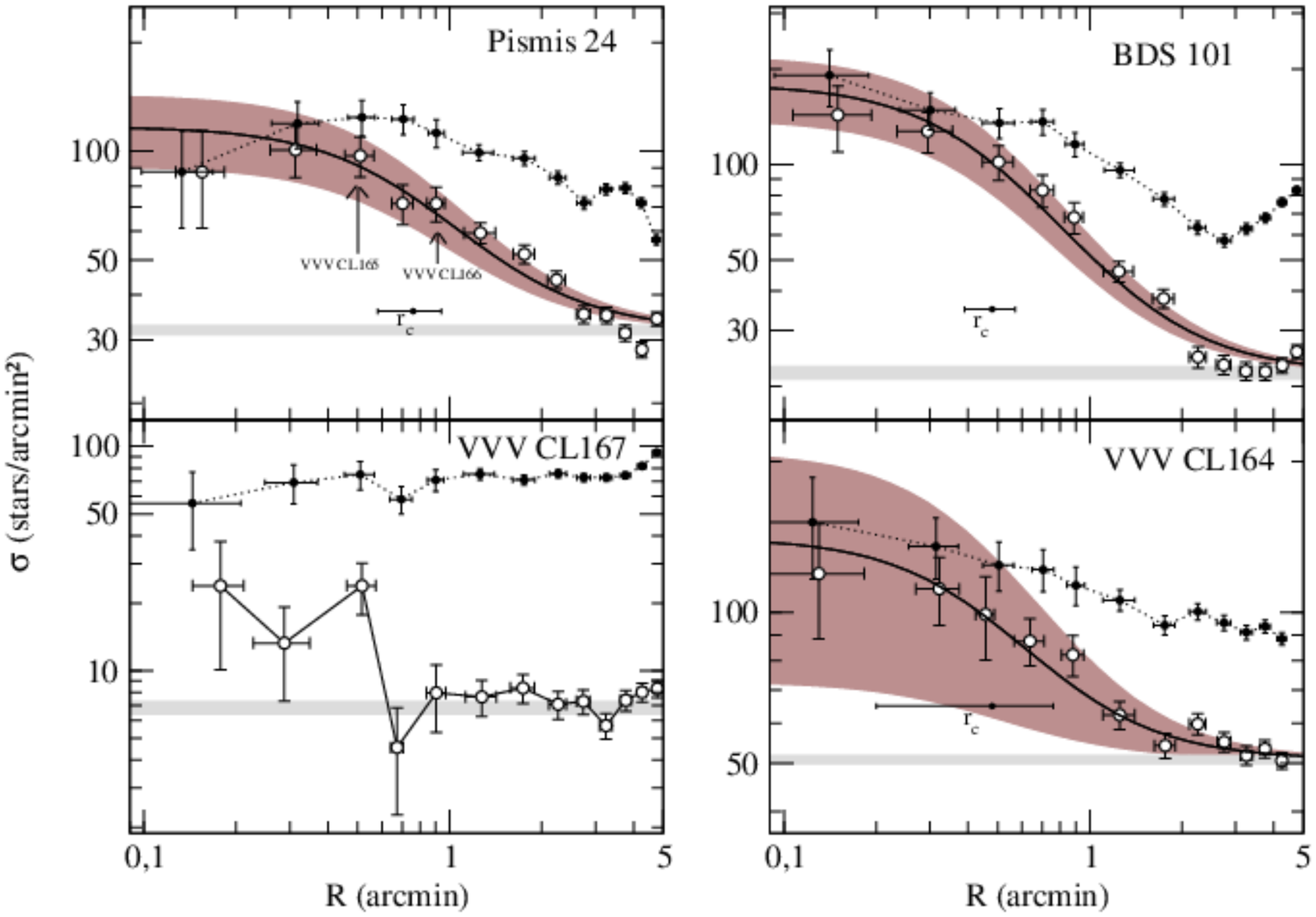}}
  \caption{Colour-magnitude filtered RDPs (empty circles). The dots represent the raw cluster profile. The empty circles show the cleaned profiles. The best King-like fit is shown, when possible. The 1$\sigma$ fit uncertainty is represented by the heavy-shaded 
  strip along the fit. The background level is the light-shaded region. The core radius (r$_c$) 
  is indicated.}
  \label{rdp}
  \end{figure}

 We use projected stellar RDPs, defined as the stellar number density around the cluster center, 
 to derive structural parameters. Noise in the RDPs is minimised with CM filters, 
 which exclude stars with colours that are not compatible with those of the cluster.
 Our group has shown that this filtering procedure considerably enhances the RDP 
 contrast relative to the comparison field, especially in crowded fields \citep{2007A&A...473..445B}. The CM filtered RDPs of the clusters are shown in Fig.~\ref{rdp}.
 
 Whenever possible, we fit a two-parameter King-like profile $\sigma$(R)=$\sigma_{bg}$+$\sigma_0/(1+(R/R_c)^2$) \citep{1966AJ.....71...64K} 
 adapted to star counts, where $\sigma_0$ and $\sigma_{bg}$ are the central and residual stellar densities, and R$_c$ 
 is the core radius. The structural $\sigma_0$ and R$_c$ parameters are derived from the fit, while $\sigma_{bg}$ is previously 
 measured in the comparison field and kept constant. The extent of the clusters (r$_{RDP}$) is defined as the radius where
  the RDP meets the field-star count level. The best-fit solutions and other results are condensed in Table \ref{tb_rdp}.  We point out that we are dealing with very young clusters (except VVV\,CL164). Thus we are not expecting relaxed systems, but interestingly the profiles even so describe the cluster RDP, such as the decontaminated profile of Pismis\,24.
  
  \begin{figure}
\centering
{\includegraphics[width=2.5in, trim= .5 .5 .5 .5, clip=true]{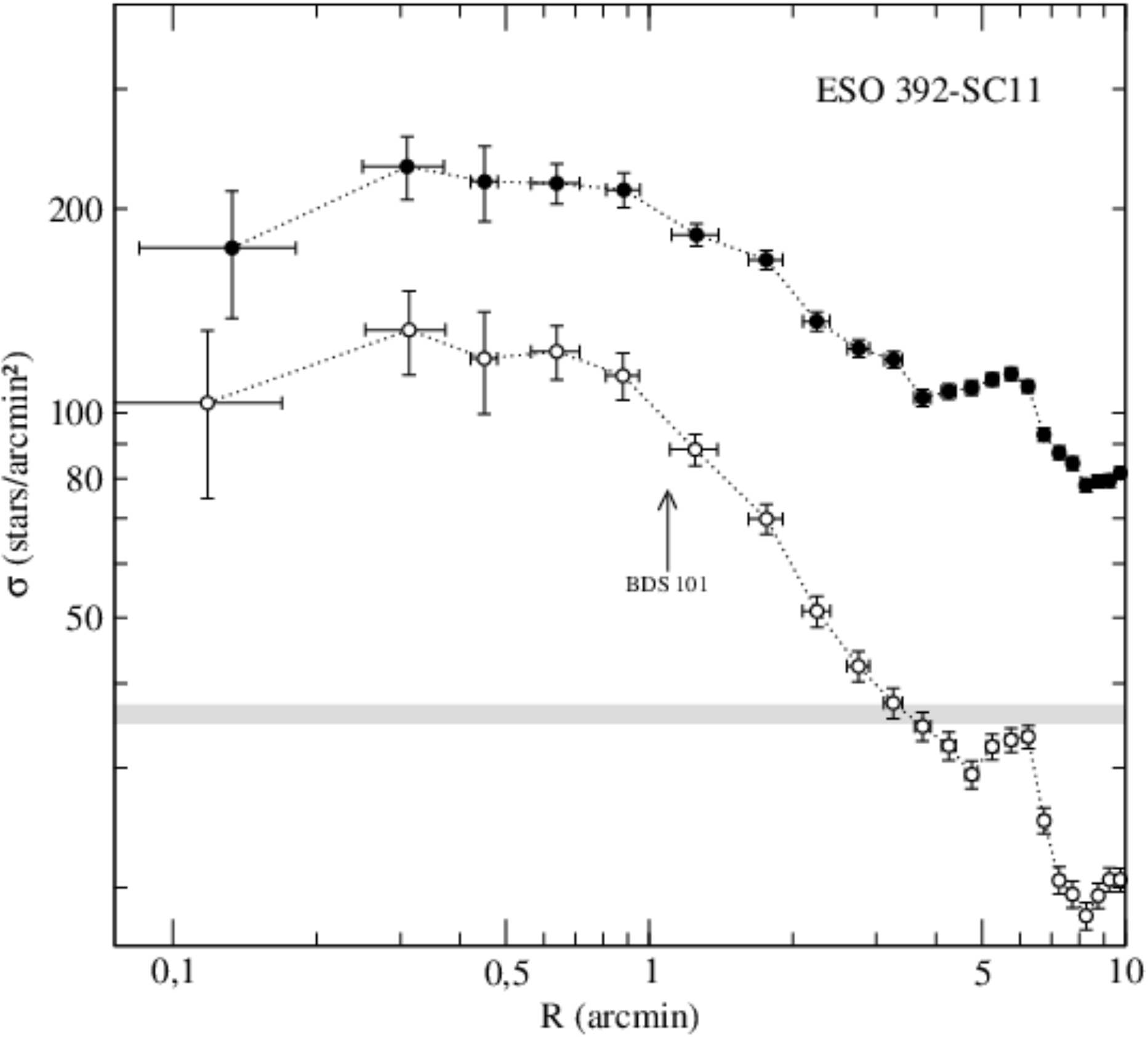}}
\caption{Same as Fig.~\ref{rdp}, for ESO\,392-SC\,11. The location of BDS\,101 in the profile is indicated.}
\label{RDP_ESO}
\end{figure}
 
 \begin{table*}
\caption{Cluster structural parameters} \label{tb_rdp}
\centering
\begin{tabular}{ccccccc}
\hline
Cluster & $\sigma_0$ & r$_c$ & R$_c$ & $\sigma_{bg}$ & r$_{RDP}$ & R$_{RDP}$\\
 & (stars/$'^2$) & ($'$) & (pc) & (stars/$'^2$) & ($'$) & (pc)\\
\hline
Pismis\,24 & 85$\pm$26 & 0.76$\pm$0.20 & 0.44$\pm$0.11 & 31.9$\pm$0.9 & 3.2$\pm$0.3 & 1.9$\pm$0.2 \\
BDS\,101 & 157$\pm$40 & 0.48$\pm$0.21 & 0.23$\pm$0.04 & 22.1$\pm$1.0& 2.3$\pm$0.2 & 1.1$\pm$0.1\\
ESO\,392-SC\,11&104$\pm$29& - & - & 36$\pm$1.1 & 3.3$\pm$0.3 &1.8$\pm$0.2\\
VVV\,CL167 & 24$\pm$14 & - & - & 6.8$\pm$0.4& 2.3$\pm$0.2 & 1.1$\pm$0.1\\
%Lima1 & - & - &- & 6.84$\pm$0.43\\
VVV\,CL164 & 90$\pm$68 & 0.48$\pm$0.28  & 0.19$\pm$0.10 & 50.8$\pm$1.0&3.3$\pm$0.3& 1.3$\pm$0.1\\
\hline
\end{tabular}
\tablefoot{Col. 1: cluster designation. Col. 2: central density.  Col. 3: core radius (arcmin). Col. 4: core radius (pc), 
assuming distances to the Sun in Table \ref{tb_par}. Col. 5: background density. Col. 6: cluster RDP radius (arcmin). Col. 7: cluster RDP radius (pc).}
\end{table*}   

The absence of a central peak in the RDP of Pismis\,24 (Fig. \ref{rdp}) is probably a consequence of the overshadowing produced by the bright stars on the surrounding faint members. The RDP appears to present two excesses, supporting the existence of subclusters there (VVV\,CL165 and VVV\,CL166). 

The irregular RDP of VVV\,CL167 could not be fitted by a King-like profile (Fig. \ref{rdp}). This suggests that 
VVV\,CL167 is a dynamically evolved cluster as a consequence of general mass loss, in particular winds from WR\,93 and other less-massive stars that remove the interstellar material in the cluster \citep{1978A&A....70...57T, 2006MNRAS.373..752G}.

ESO\,392-SC\,11 has a peculiar profile (Fig.~\ref{RDP_ESO}). We could not fit any King-profile. The superimposed contribution of the compact cluster BDS\,101 could not be subtracted.   

\section{Object structure}
\label{structure}

We use structural parameters from Sect.~\ref{rdp_sect} to investigate possible cluster evolutionary effects \citep{2011MNRAS.410L...6G, Saurin}.
Diagnostic diagrams for the dynamical evolution of star clusters
have been built by our group in a series of papers \citep[e.g.][]{2009MNRAS.394.2127B, 2008A&A...489.1129B, 2010A&A...516A..81B}.

\begin{figure}
  \centering
  {\includegraphics[width=2.5in, clip=true]{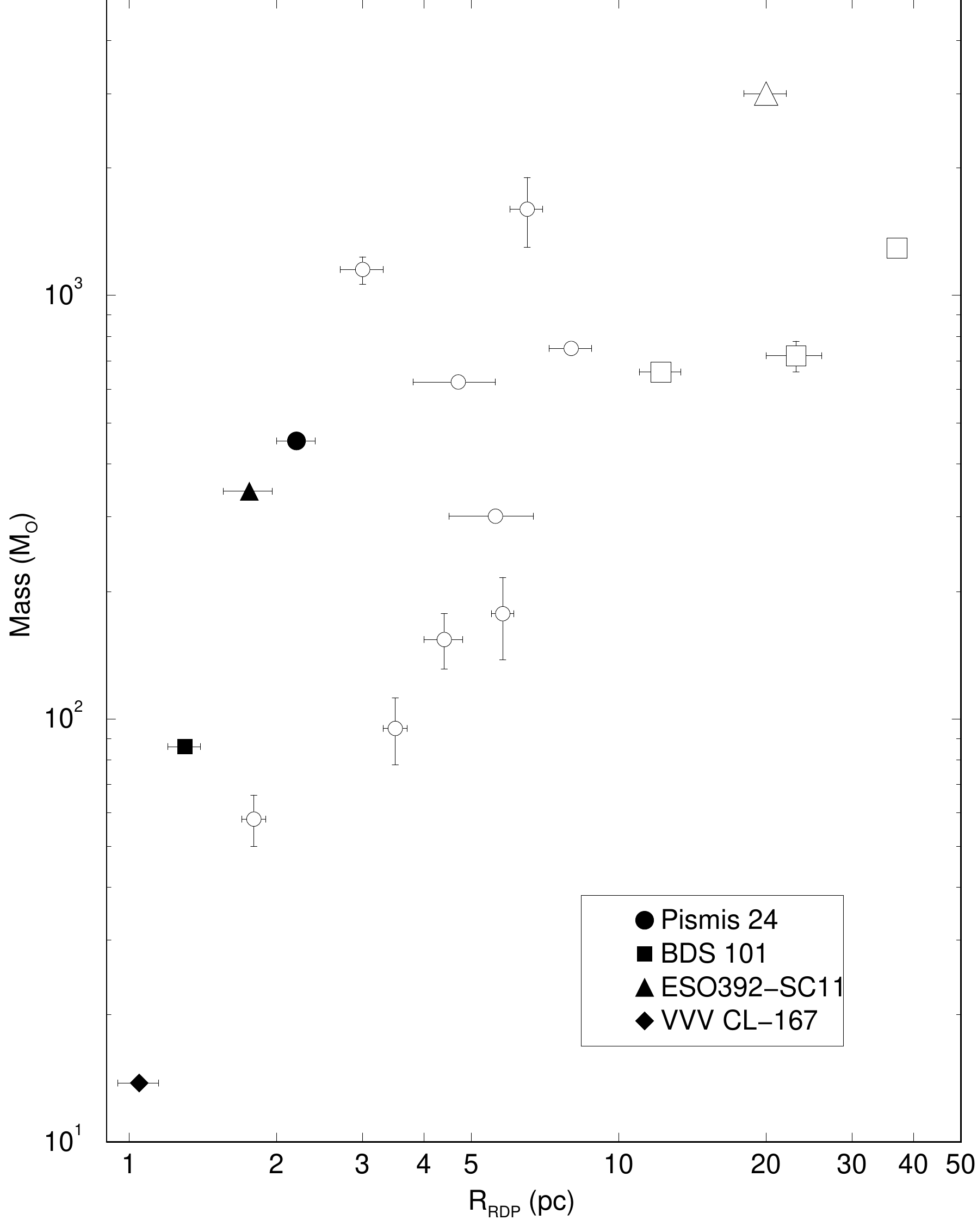}}
  \caption{Cluster or association mass $\times$ extent RDP radius. Open circles: comparison cluster sample; Open squares: associations or evolving associations; Open triangle: the LMC association LH\,95. The present sample is indicated.}
  \label{diag_mass}
  \end{figure}

   In Fig.~\ref{diag_mass} we show the estimated stellar mass compared to the extent RDP radius.
For comparison purposes, we include a sample of nearby young clusters studied
by our group following similar methods. The clusters are: NGC\,2244 and
NGC\,2239 \citep{2009MNRAS.394.2127B}, NGC\,6611 \citep{2006A&A...445..567B}, NGC\,6823 \citep{2008A&A...489.1129B}, vdB\,92 \citep{2010A&A...516A..81B}, Pismis\,5, vdB\,80, NGC\,1931 and BDSB\,96 \citep{2009MNRAS.397.1915B}. We also include the large and populous associations
Collinder\,197 \citep{2010A&A...516A..81B}, Bochum\,1 \citep{2008A&A...489.1129B}, and Trumpler\,37 \citep{Saurin}. In addition,
we show LH\,95, a massive LMC association studied with HST \citep{2012MNRAS.422.3356D}.
The general behaviour of the clusters and associations appears to present
a flattening effect, in the sense that associations tend to inflate while
keeping large masses. The clusters in NGC\,6357 present low to intermediate
integrated stellar masses, but they are systematically smaller than the
comparison clusters. Since they are located in the Sagittarius Arm, they
are possibily affected by tidal effects, even at the phase of cloud fragmentation.

\begin{figure}
  \centering
  {\includegraphics[width=2.5in, clip=true]{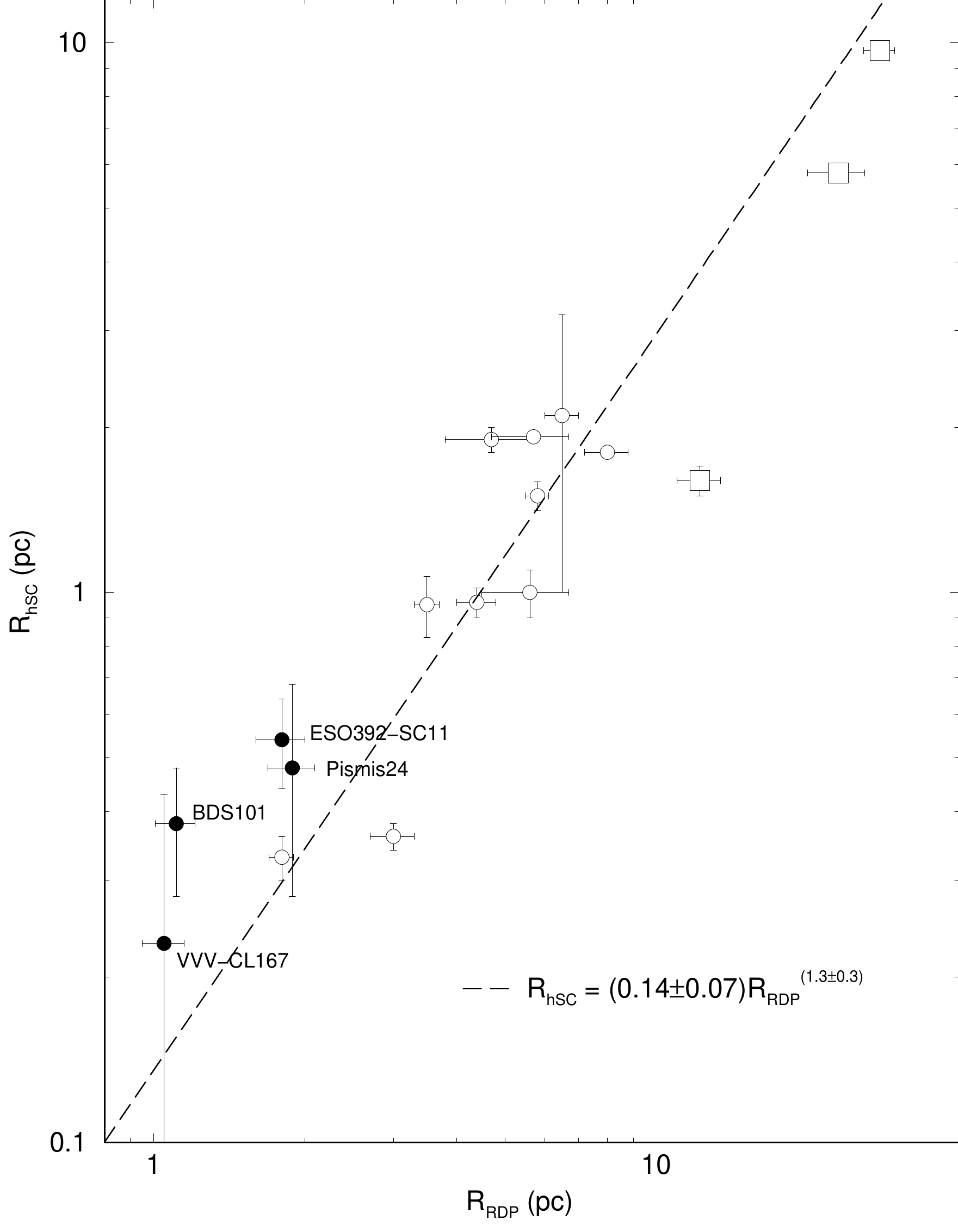}}
  \caption{The distribution of the half-starcount radius $\times$ extent radius. Symbols as in  Fig.~\ref{diag_mass}.}
  \label{diag_rdp}
  \end{figure}

Another way of investigating this issue is by comparing the cluster radius (R$_{RDP}$)
with the half-starcount radius (R$_{hSC}$). The latter is equivalent to the half-light
radius usually used in star clusters studies in the outer parts of the Galaxy \citep[e.g.][]{2009MNRAS.397.1748B}.
The results are shown in Fig.~\ref{diag_rdp}, where again we use the same comparison clusters
and associations (except LH\,95) as in Fig.~\ref{diag_mass}. The comparison clusters follow 
a power-law scaling relation for R$_{RDP}$ and R$_{hSC}$. Again, the associations Collinder\,197,
Bochum\,1 and Trumpler\,37 appear to be inflated at the outskirts, when compared
to the cluster scaling relation. The clusters in NGC\,6357 are quite small when compared to the
young reference clusters (Fig.~\ref{diag_mass}). Note that ESO\,392-SC11 (Fig.~\ref{RDP_ESO}) lies 
in a similar locus as Pismis\,24 (Fig.~\ref{diag_mass}), thus implying a cluster behaviour.

\section{Discussion and Conclusions}
\label{concl}

We studied the NGC\,6357 star-forming complex with VVV photometry. The main focus of the present study are seven stellar clusters in the area. Pismis\,24 is a relatively populous EC, with a developed MS and PMS. BDS\,101 is compact with a CMD dominated by PMS stars. The EC VVV\,CL167 includes a WR star and is significantly older than Pismis\,24 and BDS\,101. VVV\,CL165 and VVV\,CL166 appear to be subclusters related to Pismis\,24. ESO392-SC11
is an interesting object, probably in a pre-association stage. 

VVV\,CL164 to VVV\,CL167 were discovered in the present paper. From the clusters belonging to the complex (Pismis\,24, BDS\,101, ESO\,392-SC\,11 and VVV\,CL167) we derived a mean distance of d$_{\odot}$=1.78$\pm$0.1 kpc to NGC\,6357. We also detected an open cluster
of intermediate age (VVV\,CL164) projected next to the complex. Evidence is found that these clusters in NGC\,6357 are small as compared to nearby embedded clusters. The NGC\,6357 complex, now with several analysed ECs, resembles more its neighbouring complex NGC\,6334 \citep{2009AJ....138..227F,Russeil_2012,Russeil_2013}. 

In the cluster mass calculations (Sect.~\ref{CMDs}) we considered stars in the decontaminated area (corresponding to the radii given in column\,6 of Table~\ref{table1}). Pismis\,24 and ESO\,392-SC\,11 have intermediate masses, while the other clusters in NGC\,6357 have low masses. Probably VVV\,CL167 is unstable and will lose additional gas mass in the coming Myrs \citep{2006MNRAS.369L...9B}.

In some aspects, the NGC\,6357 complex emulates a star-forming dwarf galaxy, in the sense that besides a rich content of interstellar features it hosts embedded clusters pointing to at least two stellar generation events.

\begin{acknowledgements}
We thank the referee for important comments and suggestions.
We acknowledge support from the Brazilian Institution CNPq.
We gratefully acknowledge use of data from the ESO Public Survey
programme ID 179.B-2002 taken with the VISTA telescope, data products
from the Cambridge Astronomical Survey Unit.
R.K.S. acknowledges support from CNPq/Brazil through projects 310636/2013-2 and 481468/2013-7.
We acknowledge use of the VizieR Catalogue Service operated at the CDS, Strasbourg,
France. This publication makes use of data products from
the Two Micron All Sky Survey, which is a joint project of
the University of Massachusetts and the Infrared Processing and Analysis
Centre/California Institute of Technology, funded by the National Aeronautics and
Space Administration and the National Science Foundation.   
\end{acknowledgements}

%%%%%%%%%%%%%%%%%%%%%%%%%%%%%%%%%%%%%%%%%%%%%%%%%%%%%%%%%%%%%%%%%%%%%%%%%%%%
%% references
\bibliographystyle{aa} %% style aa.bst
\bibliography{ngc6357}

\begin{thebibliography}{73}
\expandafter\ifx\csname natexlab\endcsname\relax\def\natexlab#1{#1}\fi

\bibitem[{{Archinal} \& {Hynes}(2003)}]{2003stcl.book.....A}
{Archinal}, B.~A. \& {Hynes}, S.~J. 2003, {Star clusters}

\bibitem[{{Bastian} \& {Goodwin}(2006)}]{2006MNRAS.369L...9B}
{Bastian}, N. \& {Goodwin}, S.~P. 2006, \mnras, 369, L9

\bibitem[{{Belokurov} {et~al.}(2009){Belokurov}, {Walker}, {Evans}, {Gilmore},
  {Irwin}, {Mateo}, {Mayer}, {Olszewski}, {Bechtold}, \&
  {Pickering}}]{2009MNRAS.397.1748B}
{Belokurov}, V., {Walker}, M.~G., {Evans}, N.~W., {et~al.} 2009, \mnras, 397,
  1748

\bibitem[{{Bica} {et~al.}(2006){Bica}, {Bonatto}, {Barbuy}, \&
  {Ortolani}}]{2006A&A...450..105B}
{Bica}, E., {Bonatto}, C., {Barbuy}, B., \& {Ortolani}, S. 2006, \aap, 450, 105

\bibitem[{{Bica} {et~al.}(2008){Bica}, {Bonatto}, \&
  {Dutra}}]{2008A&A...489.1129B}
{Bica}, E., {Bonatto}, C., \& {Dutra}, C.~M. 2008, \aap, 489, 1129

\bibitem[{{Bica} {et~al.}(2003){Bica}, {Dutra}, {Soares}, \&
  {Barbuy}}]{2003A&A...404..223B}
{Bica}, E., {Dutra}, C.~M., {Soares}, J., \& {Barbuy}, B. 2003, \aap, 404, 223

\bibitem[{{Bica} \& {Schmitt}(1995)}]{1995ApJS..101...41B}
{Bica}, E.~L.~D. \& {Schmitt}, H.~R. 1995, \apjs, 101, 41

\bibitem[{{Bica} {et~al.}(1999){Bica}, {Schmitt}, {Dutra}, \&
  {Oliveira}}]{1999AJ....117..238B}
{Bica}, E.~L.~D., {Schmitt}, H.~R., {Dutra}, C.~M., \& {Oliveira}, H.~L. 1999,
  \aj, 117, 238

\bibitem[{{Bohigas} {et~al.}(2004){Bohigas}, {Tapia}, {Roth}, \&
  {Ruiz}}]{Bohigas_2004}
{Bohigas}, J., {Tapia}, M., {Roth}, M., \& {Ruiz}, M.~T. 2004, \aj, 127, 2826

\bibitem[{{Bonatto} \& {Bica}(2003)}]{2003A&A...405..525B}
{Bonatto}, C. \& {Bica}, E. 2003, \aap, 405, 525

\bibitem[{{Bonatto} \& {Bica}(2007{\natexlab{a}})}]{2007A&A...473..445B}
{Bonatto}, C. \& {Bica}, E. 2007{\natexlab{a}}, \aap, 473, 445

\bibitem[{{Bonatto} \& {Bica}(2007{\natexlab{b}})}]{2007MNRAS.377.1301B}
{Bonatto}, C. \& {Bica}, E. 2007{\natexlab{b}}, \mnras, 377, 1301

\bibitem[{{Bonatto} \& {Bica}(2009{\natexlab{a}})}]{2009MNRAS.392..483B}
{Bonatto}, C. \& {Bica}, E. 2009{\natexlab{a}}, \mnras, 392, 483

\bibitem[{{Bonatto} \& {Bica}(2009{\natexlab{b}})}]{2009MNRAS.394.2127B}
{Bonatto}, C. \& {Bica}, E. 2009{\natexlab{b}}, \mnras, 394, 2127

\bibitem[{{Bonatto} \& {Bica}(2009{\natexlab{c}})}]{2009MNRAS.397.1915B}
{Bonatto}, C. \& {Bica}, E. 2009{\natexlab{c}}, \mnras, 397, 1915

\bibitem[{{Bonatto} \& {Bica}(2010)}]{2010A&A...516A..81B}
{Bonatto}, C. \& {Bica}, E. 2010, \aap, 516, A81

\bibitem[{{Bonatto} {et~al.}(2012{\natexlab{a}}){Bonatto}, {Bica}, \&
  {Lima}}]{2012MNRAS.420..352B}
{Bonatto}, C., {Bica}, E., \& {Lima}, E.~F. 2012{\natexlab{a}}, \mnras, 420,
  352

\bibitem[{{Bonatto} {et~al.}(2006{\natexlab{a}}){Bonatto}, {Bica}, {Ortolani},
  \& {Barbuy}}]{2006A&A...453..121B}
{Bonatto}, C., {Bica}, E., {Ortolani}, S., \& {Barbuy}, B. 2006{\natexlab{a}},
  \aap, 453, 121

\bibitem[{{Bonatto} {et~al.}(2012{\natexlab{b}}){Bonatto}, {Lima}, \&
  {Bica}}]{2012A&A...540A.137B}
{Bonatto}, C., {Lima}, E.~F., \& {Bica}, E. 2012{\natexlab{b}}, \aap, 540, A137

\bibitem[{{Bonatto} {et~al.}(2006{\natexlab{b}}){Bonatto}, {Santos}, \&
  {Bica}}]{2006A&A...445..567B}
{Bonatto}, C., {Santos}, Jr., J.~F.~C., \& {Bica}, E. 2006{\natexlab{b}}, \aap,
  445, 567

\bibitem[{{Borissova} {et~al.}(2014){Borissova}, {Chene}, {Ramirez Alegrìa},
  {Saurabh Sharma}, {Clarke}, {Kurtev}, {Negueruela}, {Marco}, {Amigo},
  {Minniti}, {Bica}, {Bonatto}, {Caletan}, {Fierro}, {Gromadzki}, {Hempel},
  {Hanson}, {Ivanov}, {Lucas}, {Majaess}, {Moni Bidin}, {Popescu}, \&
  {Saito}}]{borissova}
{Borissova}, J., {Chene}, A.~N., {Ramirez Alegrìa}, S., {et~al.} 2014,
  submitted for publication

\bibitem[{{Bressan} {et~al.}(2012){Bressan}, {Marigo}, {Girardi}, {Salasnich},
  {Dal Cero}, {Rubele}, \& {Nanni}}]{2012MNRAS.427..127B}
{Bressan}, A., {Marigo}, P., {Girardi}, L., {et~al.} 2012, \mnras, 427, 127

\bibitem[{{Camargo} {et~al.}(2011){Camargo}, {Bonatto}, \&
  {Bica}}]{2011MNRAS.416.1522C}
{Camargo}, D., {Bonatto}, C., \& {Bica}, E. 2011, \mnras, 416, 1522

\bibitem[{{Cappa} {et~al.}(2011){Cappa}, {Barb{\'a}}, {Duronea}, {Vasquez},
  {Arnal}, {Goss}, \& {Fern{\'a}ndez Laj{\'u}s}}]{Cappa_2011}
{Cappa}, C.~E., {Barb{\'a}}, R., {Duronea}, N.~U., {et~al.} 2011, \mnras, 415,
  2844

\bibitem[{{Carvalho} {et~al.}(2008){Carvalho}, {Saurin}, {Bica}, {Bonatto}, \&
  {Schmidt}}]{Carvalho_2008}
{Carvalho}, L., {Saurin}, T.~A., {Bica}, E., {Bonatto}, C., \& {Schmidt}, A.~A.
  2008, \aap, 485, 71

\bibitem[{{Chen{\'e}} {et~al.}(2012){Chen{\'e}}, {Borissova}, {Clarke},
  {Bonatto}, {Majaess}, {Moni Bidin}, {Sale}, {Mauro}, {Kurtev}, {Baume},
  {Feinstein}, {Ivanov}, {Geisler}, {Catelan}, {Minniti}, {Lucas}, {de Grijs},
  \& {Kumar}}]{2012A&A...545A..54C}
{Chen{\'e}}, A.-N., {Borissova}, J., {Clarke}, J.~R.~A., {et~al.} 2012, \aap,
  545, A54

\bibitem[{{Conti} \& {Vacca}(1990)}]{Conti_1990}
{Conti}, P.~S. \& {Vacca}, W.~D. 1990, \aj, 100, 431

\bibitem[{{Da Rio} {et~al.}(2012){Da Rio}, {Gouliermis}, {Rochau}, {Pasquali},
  {Setiawan}, \& {De Marchi}}]{2012MNRAS.422.3356D}
{Da Rio}, N., {Gouliermis}, D.~A., {Rochau}, B., {et~al.} 2012, \mnras, 422,
  3356

\bibitem[{{Damke} {et~al.}(2006){Damke}, {Barb{\'a}}, \&
  {Morrell}}]{2006RMxAC..26..180D}
{Damke}, G., {Barb{\'a}}, R., \& {Morrell}, N.~I. 2006, in Revista Mexicana de
  Astronomia y Astrofisica, vol. 27, Vol.~26, Revista Mexicana de Astronomia y
  Astrofisica Conference Series, 180

\bibitem[{{Dias} {et~al.}(2002){Dias}, {Alessi}, {Moitinho}, \&
  {L{\'e}pine}}]{2002A&A...389..871D}
{Dias}, W.~S., {Alessi}, B.~S., {Moitinho}, A., \& {L{\'e}pine}, J.~R.~D. 2002,
  \aap, 389, 871

\bibitem[{{Fang} {et~al.}(2012){Fang}, {van Boekel}, {King}, {Henning},
  {Bouwman}, {Doi}, {Okamoto}, {Roccatagliata}, \&
  {Sicilia-Aguilar}}]{Fang_2012}
{Fang}, M., {van Boekel}, R., {King}, R.~R., {et~al.} 2012, \aap, 539, A119

\bibitem[{{Feigelson} {et~al.}(2009){Feigelson}, {Martin}, {McNeill}, {Broos},
  \& {Garmire}}]{2009AJ....138..227F}
{Feigelson}, E.~D., {Martin}, A.~L., {McNeill}, C.~J., {Broos}, P.~S., \&
  {Garmire}, G.~P. 2009, \aj, 138, 227

\bibitem[{{Felli} {et~al.}(1990){Felli}, {Persi}, {Roth}, {Tapia},
  {Ferrari-Toniolo}, \& {Cervelli}}]{Felli_1990}
{Felli}, M., {Persi}, P., {Roth}, M., {et~al.} 1990, \aap, 232, 477

\bibitem[{{Friel}(1995)}]{1995ARA&A..33..381F}
{Friel}, E.~D. 1995, \araa, 33, 381

\bibitem[{{Gieles} \& {Portegies Zwart}(2011)}]{2011MNRAS.410L...6G}
{Gieles}, M. \& {Portegies Zwart}, S.~F. 2011, \mnras, 410, L6

\bibitem[{{Goodwin} \& {Bastian}(2006)}]{2006MNRAS.373..752G}
{Goodwin}, S.~P. \& {Bastian}, N. 2006, \mnras, 373, 752

\bibitem[{{Gouliermis} {et~al.}(2007){Gouliermis}, {Henning}, {Brandner},
  {Dolphin}, {Rosa}, \& {Brandl}}]{2007ApJ...665L..27G}
{Gouliermis}, D.~A., {Henning}, T., {Brandner}, W., {et~al.} 2007, \apjl, 665,
  L27

\bibitem[{{Hodge}(1985)}]{1985PASP...97..530H}
{Hodge}, P. 1985, \pasp, 97, 530

\bibitem[{{Irwin} {et~al.}(2004){Irwin}, {Lewis}, {Hodgkin}, {Bunclark},
  {Evans}, {McMahon}, {Emerson}, {Stewart}, \& {Beard}}]{Irwin_2004}
{Irwin}, M.~J., {Lewis}, J., {Hodgkin}, S., {et~al.} 2004, in Society of
  Photo-Optical Instrumentation Engineers (SPIE) Conference Series, Vol. 5493,
  Optimizing Scientific Return for Astronomy through Information Technologies,
  ed. P.~J. {Quinn} \& A.~{Bridger}, 411--422

\bibitem[{{King}(1966)}]{1966AJ.....71...64K}
{King}, I.~R. 1966, \aj, 71, 64

\bibitem[{{Kroupa}(2001)}]{2001MNRAS.322..231K}
{Kroupa}, P. 2001, \mnras, 322, 231

\bibitem[{{Lada} \& {Lada}(2003)}]{2003ARA&A..41...57L}
{Lada}, C.~J. \& {Lada}, E.~A. 2003, \araa, 41, 57

\bibitem[{{Lauberts}(1982)}]{1982euse.book.....L}
{Lauberts}, A. 1982, {ESO/Uppsala survey of the ESO(B) atlas}

\bibitem[{{Leisawitz} {et~al.}(1989){Leisawitz}, {Bash}, \&
  {Thaddeus}}]{1989ApJS...70..731L}
{Leisawitz}, D., {Bash}, F.~N., \& {Thaddeus}, P. 1989, \apjs, 70, 731

\bibitem[{{Lortet} {et~al.}(1984){Lortet}, {Testor}, \&
  {Niemela}}]{Lortet_1984}
{Lortet}, M.~C., {Testor}, G., \& {Niemela}, V. 1984, \aap, 140, 24

\bibitem[{{Lucke} \& {Hodge}(1970)}]{1970AJ.....75..171L}
{Lucke}, P.~B. \& {Hodge}, P.~W. 1970, \aj, 75, 171

\bibitem[{{Ma{\'{\i}}z Apell{\'a}niz} {et~al.}(2007){Ma{\'{\i}}z
  Apell{\'a}niz}, {Walborn}, {Morrell}, {Niemela}, \&
  {Nelan}}]{2007ApJ...660.1480M}
{Ma{\'{\i}}z Apell{\'a}niz}, J., {Walborn}, N.~R., {Morrell}, N.~I., {Niemela},
  V.~S., \& {Nelan}, E.~P. 2007, \apj, 660, 1480

\bibitem[{{Massey} {et~al.}(2001){Massey}, {DeGioia-Eastwood}, \&
  {Waterhouse}}]{Massey_2001}
{Massey}, P., {DeGioia-Eastwood}, K., \& {Waterhouse}, E. 2001, \aj, 121, 1050

\bibitem[{{McMillan} {et~al.}(2007){McMillan}, {Vesperini}, \& {Portegies
  Zwart}}]{2007ApJ...655L..45M}
{McMillan}, S.~L.~W., {Vesperini}, E., \& {Portegies Zwart}, S.~F. 2007, \apjl,
  655, L45

\bibitem[{{Megeath} {et~al.}(2004){Megeath}, {Allen}, {Gutermuth}, {Pipher},
  {Myers}, {Calvet}, {Hartmann}, {Muzerolle}, \& {Fazio}}]{2004ApJS..154..367M}
{Megeath}, S.~T., {Allen}, L.~E., {Gutermuth}, R.~A., {et~al.} 2004, \apjs,
  154, 367

\bibitem[{{Meynet} \& {Maeder}(2005)}]{2005A&A...429..581M}
{Meynet}, G. \& {Maeder}, A. 2005, \aap, 429, 581

\bibitem[{{Minniti} {et~al.}(2010){Minniti}, {Lucas}, {Emerson}, {Saito},
  {Hempel}, {Pietrukowicz}, {Ahumada}, {Alonso}, {Alonso-Garcia}, {Arias},
  {Bandyopadhyay}, {Barb{\'a}}, {Barbuy}, {Bedin}, {Bica}, {Borissova},
  {Bronfman}, {Carraro}, {Catelan}, {Clari{\'a}}, {Cross}, {de Grijs},
  {D{\'e}k{\'a}ny}, {Drew}, {Fari{\~n}a}, {Feinstein}, {Fern{\'a}ndez
  Laj{\'u}s}, {Gamen}, {Geisler}, {Gieren}, {Goldman}, {Gonzalez}, {Gunthardt},
  {Gurovich}, {Hambly}, {Irwin}, {Ivanov}, {Jord{\'a}n}, {Kerins}, {Kinemuchi},
  {Kurtev}, {L{\'o}pez-Corredoira}, {Maccarone}, {Masetti}, {Merlo},
  {Messineo}, {Mirabel}, {Monaco}, {Morelli}, {Padilla}, {Palma}, {Parisi},
  {Pignata}, {Rejkuba}, {Roman-Lopes}, {Sale}, {Schreiber}, {Schr{\"o}der},
  {Smith}, {Sodr{\'e}}, {Soto}, {Tamura}, {Tappert}, {Thompson}, {Toledo},
  {Zoccali}, \& {Pietrzynski}}]{Minniti_2010}
{Minniti}, D., {Lucas}, P.~W., {Emerson}, J.~P., {et~al.} 2010, \na, 15, 433

\bibitem[{{Moffat} \& {Vogt}(1973)}]{Moffat_1973}
{Moffat}, A.~F.~J. \& {Vogt}, N. 1973, \aaps, 10, 135

\bibitem[{{Mois{\'e}s} {et~al.}(2011){Mois{\'e}s}, {Damineli}, {Figuer{\^e}do},
  {Blum}, {Conti}, \& {Barbosa}}]{2011MNRAS.411..705M}
{Mois{\'e}s}, A.~P., {Damineli}, A., {Figuer{\^e}do}, E., {et~al.} 2011,
  \mnras, 411, 705

\bibitem[{{Naylor} \& {Jeffries}(2006)}]{2006MNRAS.373.1251N}
{Naylor}, T. \& {Jeffries}, R.~D. 2006, \mnras, 373, 1251

\bibitem[{{Neckel}(1978)}]{Neckel_1978}
{Neckel}, T. 1978, \aap, 69, 51

\bibitem[{{Neckel}(1984)}]{Neckel_1984}
{Neckel}, T. 1984, \aap, 137, 58

\bibitem[{{Pang} {et~al.}(2013){Pang}, {Grebel}, {Allison}, {Goodwin},
  {Altmann}, {Harbeck}, {Moffat}, \& {Drissen}}]{2013ApJ...764...73P}
{Pang}, X., {Grebel}, E.~K., {Allison}, R.~J., {et~al.} 2013, \apj, 764, 73

\bibitem[{{Persi} {et~al.}(1986){Persi}, {Ferrari-Toniolo}, {Roth}, \&
  {Tapia}}]{1986A&A...170...97P}
{Persi}, P., {Ferrari-Toniolo}, M., {Roth}, M., \& {Tapia}, M. 1986, \aap, 170,
  97

\bibitem[{{Rahman} {et~al.}(2011{\natexlab{a}}){Rahman}, {Matzner}, \&
  {Moon}}]{2011ApJ...728L..37R}
{Rahman}, M., {Matzner}, C., \& {Moon}, D.-S. 2011{\natexlab{a}}, \apjl, 728,
  L37

\bibitem[{{Rahman} {et~al.}(2011{\natexlab{b}}){Rahman}, {Moon}, \&
  {Matzner}}]{2011ApJ...743L..28R}
{Rahman}, M., {Moon}, D.-S., \& {Matzner}, C.~D. 2011{\natexlab{b}}, \apjl,
  743, L28

\bibitem[{{Russeil} {et~al.}(2013){Russeil}, {Schneider}, {Anderson},
  {Zavagno}, {Molinari}, {Persi}, {Bontemps}, {Motte}, {Ossenkopf},
  {Andr{\'e}}, {Arzoumanian}, {Bernard}, {Deharveng}, {Didelon}, {Di
  Francesco}, {Elia}, {Hennemann}, {Hill}, {K{\"o}nyves}, {Li}, {Martin},
  {Nguyen Luong}, {Peretto}, {Pezzuto}, {Polychroni}, {Roussel}, {Rygl},
  {Spinoglio}, {Testi}, {Tig{\'e}}, {Vavrek}, {Ward-Thompson}, \&
  {White}}]{Russeil_2013}
{Russeil}, D., {Schneider}, N., {Anderson}, L.~D., {et~al.} 2013, \aap, 554,
  A42

\bibitem[{{Russeil} {et~al.}(2012){Russeil}, {Zavagno}, {Adami}, {Anderson},
  {Bontemps}, {Motte}, {Rodon}, {Schneider}, {Ilmane}, \&
  {Murphy}}]{Russeil_2012}
{Russeil}, D., {Zavagno}, A., {Adami}, C., {et~al.} 2012, \aap, 538, A142

\bibitem[{{Russeil} {et~al.}(2010){Russeil}, {Zavagno}, {Motte}, {Schneider},
  {Bontemps}, \& {Walsh}}]{Russeil_2010}
{Russeil}, D., {Zavagno}, A., {Motte}, F., {et~al.} 2010, \aap, 515, A55

\bibitem[{{Saito} {et~al.}(2012){Saito}, {Hempel}, {Minniti}, {Lucas},
  {Rejkuba}, {Toledo}, {Gonzalez}, {Alonso-Garc{\'{\i}}a}, {Irwin},
  {Gonzalez-Solares}, {Hodgkin}, {Lewis}, {Cross}, {Ivanov}, {Kerins},
  {Emerson}, {Soto}, {Am{\^o}res}, {Gurovich}, {D{\'e}k{\'a}ny}, {Angeloni},
  {Beamin}, {Catelan}, {Padilla}, {Zoccali}, {Pietrukowicz}, {Moni Bidin},
  {Mauro}, {Geisler}, {Folkes}, {Sale}, {Borissova}, {Kurtev}, {Ahumada},
  {Alonso}, {Adamson}, {Arias}, {Bandyopadhyay}, {Barb{\'a}}, {Barbuy},
  {Baume}, {Bedin}, {Bellini}, {Benjamin}, {Bica}, {Bonatto}, {Bronfman},
  {Carraro}, {Chen{\`e}}, {Clari{\'a}}, {Clarke}, {Contreras}, {Corvill{\'o}n},
  {de Grijs}, {Dias}, {Drew}, {Fari{\~n}a}, {Feinstein},
  {Fern{\'a}ndez-Laj{\'u}s}, {Gamen}, {Gieren}, {Goldman},
  {Gonz{\'a}lez-Fern{\'a}ndez}, {Grand}, {Gunthardt}, {Hambly}, {Hanson},
  {He{\l}miniak}, {Hoare}, {Huckvale}, {Jord{\'a}n}, {Kinemuchi}, {Longmore},
  {L{\'o}pez-Corredoira}, {Maccarone}, {Majaess}, {Mart{\'{\i}}n}, {Masetti},
  {Mennickent}, {Mirabel}, {Monaco}, {Morelli}, {Motta}, {Palma}, {Parisi},
  {Parker}, {Pe{\~n}aloza}, {Pietrzy{\'n}ski}, {Pignata}, {Popescu}, {Read},
  {Rojas}, {Roman-Lopes}, {Ruiz}, {Saviane}, {Schreiber}, {Schr{\"o}der},
  {Sharma}, {Smith}, {Sodr{\'e}}, {Stead}, {Stephens}, {Tamura}, {Tappert},
  {Thompson}, {Valenti}, {Vanzi}, {Walton}, {Weidmann}, \&
  {Zijlstra}}]{Saito_2012}
{Saito}, R.~K., {Hempel}, M., {Minniti}, D., {et~al.} 2012, \aap, 537, A107

\bibitem[{{Saurin} {et~al.}(2010){Saurin}, {Bica}, \&
  {Bonatto}}]{2010MNRAS.407..133S}
{Saurin}, T.~A., {Bica}, E., \& {Bonatto}, C. 2010, \mnras, 407, 133

\bibitem[{{Saurin} {et~al.}(2012){Saurin}, {Bica}, \& {Bonatto}}]{Saurin}
{Saurin}, T.~A., {Bica}, E., \& {Bonatto}, C. 2012, \mnras, 421, 3206

\bibitem[{{Skrutskie} {et~al.}(2006){Skrutskie}, {Cutri}, {Stiening},
  {Weinberg}, {Schneider}, {Carpenter}, {Beichman}, {Capps}, {Chester},
  {Elias}, {Huchra}, {Liebert}, {Lonsdale}, {Monet}, {Price}, {Seitzer},
  {Jarrett}, {Kirkpatrick}, {Gizis}, {Howard}, {Evans}, {Fowler}, {Fullmer},
  {Hurt}, {Light}, {Kopan}, {Marsh}, {McCallon}, {Tam}, {Van Dyk}, \&
  {Wheelock}}]{Skrutskie_2006}
{Skrutskie}, M.~F., {Cutri}, R.~M., {Stiening}, R., {et~al.} 2006, \aj, 131,
  1163

\bibitem[{{Tutukov}(1978)}]{1978A&A....70...57T}
{Tutukov}, A.~V. 1978, \aap, 70, 57

\bibitem[{{van der Hucht}(2001)}]{Van_der_Hucht_01}
{van der Hucht}, K.~A. 2001, \nar, 45, 135

\bibitem[{{Walborn} {et~al.}(2002){Walborn}, {Howarth}, {Lennon}, {Massey},
  {Oey}, {Moffat}, {Skalkowski}, {Morrell}, {Drissen}, \&
  {Parker}}]{Walborn_2002}
{Walborn}, N.~R., {Howarth}, I.~D., {Lennon}, D.~J., {et~al.} 2002, \aj, 123,
  2754

\bibitem[{{Wang} {et~al.}(2007){Wang}, {Townsley}, {Feigelson}, {Getman},
  {Broos}, {Garmire}, \& {Tsujimoto}}]{Wang_2007}
{Wang}, J., {Townsley}, L.~K., {Feigelson}, E.~D., {et~al.} 2007, \apjs, 168,
  100

\bibitem[{{Wilson} {et~al.}(1970){Wilson}, {Mezger}, {Gardner}, \&
  {Milne}}]{Wilson_1970}
{Wilson}, T.~L., {Mezger}, P.~G., {Gardner}, F.~F., \& {Milne}, D.~K. 1970,
  \aap, 6, 364

\end{thebibliography}

\end{document}